\documentclass[11pt]{article}
\usepackage{geometry}
\usepackage{Shorthands}

\title{Adversarial Tradeoffs in Robust State Estimation}

\author{Thomas T.C.K.\ Zhang\thanks{Equal contribution} , Bruce D.\ Lee\footnotemark[1] , Hamed Hassani, and Nikolai Matni}

\date{Department of Electrical and Systems Engineering, University of Pennsylvania}

\begin{document}

\maketitle

\vspace{-0.5cm}
\begin{abstract}

    Adversarially robust training has been shown to reduce the susceptibility of learned models to targeted input data perturbations.  However, it has also been observed that such adversarially robust models suffer a degradation in accuracy when applied to unperturbed data sets, leading to a robustness-accuracy tradeoff. Inspired by recent progress in the adversarial machine learning literature which characterize such tradeoffs in simple settings, we develop tools to quantitatively study the performance-robustness tradeoff between nominal and robust state estimation. In particular, we define and analyze a novel \emph{adversarially robust Kalman Filtering problem.} We show that in contrast to most problem instances in adversarial machine learning, we can precisely derive the adversarial perturbation in the Kalman Filtering setting. We provide an algorithm to find this perturbation given data realizations, and develop upper and lower bounds on the adversarial state estimation error in terms of the standard (non-adversarial) estimation error and the spectral properties of the resulting observer. Through these results, we show a natural connection between a filter's robustness to adversarial perturbation and underlying control theoretic properties of the system being observed, namely the spectral properties of its observability gramian.

\end{abstract}

%

\section{Introduction}

It has been demonstrated across various application areas that contemporary learning-based models, despite their impressive nominal performance, can be extremely susceptible to small, adversarially designed input perturbations \citep{carlini2016defensive, goodfellow2014explaining, szegedy2013intriguing, huang2017adversarial}.  In order to mitigate the effects of such attacks, various adversarially robust training algorithms \citep{carlini2016defensive, carlini2017towards, madry2017towards, xie2020smooth, deka2020dynamically} have been developed.  However, it was soon noticed that while adversarial training could be used to improve model robustness, it often comes with a corresponding decrease in accuracy on nominal (unperturbed) data.  Further, various simplified theoretical models \citep{tsipras2018robustness,zhang2019theoretically,nakkiran2019adversarial,raghunathan2019adversarial,chen2020more, javanmard2020precise}, have been used to explain this phenomena, and to argue that such \emph{robustness-accuracy tradeoffs} are unavoidable.

In this paper, we extend the study of such robustness-accuracy tradeoffs to data generated by a dynamical system via the setting of \emph{adversarially robust Kalman Filtering}. Adversarial robustness is an appealing model of robust filtering, as it captures measurement disturbances that are composed of both stochastic and worst-case components. Motivated by applications of adversarial robustness in the reinforcement literature \citep{lutter2021robust,pinto2017robust,mandlekar2017adversarially}, we provide the first theoretical analysis robustness-accuracy tradeoffs for state estimation of a dynamical system, and in doing so establish connections to natural control theoretic properties of the underlying system.

Our specific contributions can be summarized as follows:
\begin{itemize}
    \item We propose a simple and computationally efficient algorithm that provably finds optimal worst-case $\ell^2$ norm-bounded adversarial perturbations of the measurements for a given observer and trajectory data. This allows us to efficiently compute and explore the Pareto-optimal robustness-accuracy tradeoff curve.
    
    
    \item We analyze the adversarially robust Kalman Filtering problem, and show that upper and lower bounds on the gap between the adversarial and standard (unperturbed) state estimation error can be controlled in terms of the spectral properties of the \emph{observability gramian} \citep{zhou1998robust} of the underlying system.
     
    \item As an intermediate step to deriving the aforementioned bounds, we bound the gap between adversarial and standard (unperturbed) risks for general linear inverse problems in terms of the spectral properties of a given linear model. We also show that our bounds are 
    tight in the one-dimensional setting, recovering the results of \cite{javanmard2020precise}, and for matrices with full column rank and orthogonal columns. These results may be of independent interest.
    
    \item We empirically demonstrate through numerical simulations that our results predict robustness-accuracy tradeoffs in Kalman Filtering as a function of underlying spectral properties of the observability gramian. 
    
\end{itemize}

The rest of this paper is organized as follows: in Section~\ref{sec: Kalman}, we pose the adversarially robust Kalman Filtering problem. In Section~\ref{sec: Kalman Bounds}, we first present tight upper and lower bounds on the adversarial risk in a general linear inverse problem, then refine these bounds for the Kalman filtering setting. The results reveal that for the setting where data is generated by a dynamical system, robustness-accuracy tradeoffs are dictated by natural control theoretic properties of the underlying system (namely, the observability gramian). In Section~\ref{sec: numerical}, we provide empirical evidence to support the trends predicted by our bounds, and demonstrate the efficacy of adversarially robust Kalman Filtering against sensor drift. We end with conclusions and a discussion of future work in Section~\ref{sec:conclusion}.


\subsection{Related Work}
Our work makes connections between adversarial robustnesss and robust estimation and control.  We now provide a brief overview of work related to ours from these areas.

\textbf{Robustness-accuracy tradeoffs:} 
We draw inspiration from recent work offering theoretical characterizations of robustness-accuracy tradeoffs: \citet{tsipras2018robustness} and \citet{zhang2019theoretically} posit that high standard accuracy is fundamentally at odds with high robust accuracy by considering classification problems, whereas \citet{nakkiran2019adversarial} suggests an alternative explanation that classifiers that are simultaneously robust and accurate are complex, and may not be contained in current function classes. However, \citet{raghunathan2019adversarial} shows that the tradeoff is not due to optimization or representation issues by showing that such tradeoffs exist even for a problem with a convex loss where the optimal predictor achieves 100\% standard and robust accuracy.  In contrast to previous work, we provide sharp and interpretable characterizations of the robustness-accuracy tradeoffs that may arise in inverse problems, albeit restricted to linear models.  Most closely related to our work, \citet{javanmard2020precise} derive a formula for the exact tradeoff between standard and robust accuracy in the linear regression setting.  We derive similar results for a matrix-valued linear inverse problem, and apply these tools to the adversarially robust Kalman Filtering problem, wherein data is generated by a dynamical system.

\textbf{Robust estimation and control:}
 Robustness in estimation and control has traditionally been studied from a worst-case induced gain perspective \citep{hassibi1999,zhou1998robust}. When perturbations are restricted to be $\ell^2$-bounded, this gives rise to  $\calH_\infty$ estimation and control problems. Although widely known, and celebrated for their applications in robust control, $\calH_\infty$-based methods are often overly conservative. This conservatism can be reduced by using mixed $\calH_2/\calH_\infty$ methods \citep{khargonekar1996}, which blend Gaussian and worst-case disturbance assumptions. While such an approach is related to the adversarially robust Kalman Filtering problem that we pose, we note that mixed $\calH_2/\calH_\infty$ decouples the worst-case and stochastic inputs during design, leading to a fundamentally different tradeoff.
 On the other hand, our method considers the stochastic and worst-case components jointly, finding the optimal filter robust to $\ell^2$-bounded disturbances \textit{given} the realizations of stochastic noise. We note the decoupling of worst-case and stochastic inputs allows mixed $\calH_2/\calH_\infty$ methods to guarantee stability under dynamic uncertainties. 
 We leave similar guarantees for the filtering problem that we pose, and further characterizing connections between traditional and adversarial robustness to future work. More recently, \cite{al2020accuracy} have considered the robustness-accuracy tradeoff in data-driven perception-based control. However, the adversary in \cite{al2020accuracy}  perturbs the covariance of the noise distribution, which is assumed to remain Gaussian, whereas our adversary additively attacks each measurement, which is more aligned to the perturbations considered in machine learning and robust control contexts. Additionally, \cite{al2020accuracy} does not quantitatively analyze the severity of tradeoffs, but rather proves the existence of tradeoffs. Our prior work \cite{lee2022performance} studies analogous tradeoffs to the ones in this paper arising in the setting of adversarially robust LQR, and bounds the severity of these tradeoffs in terms of the spectral properties of the controllability gramian.
\section{Adversarially Robust Kalman Filtering}
\label{sec: Kalman}
\sloppy

We consider a modification to the standard Kalman Filtering problem to incorporate adversarial robustness.  We then bound the inflation of the state estimation error caused by the adversary, with the goal of relating control theoretic properties of the underlying linear dynamical system to the robustness-accuracy tradeoffs that it induces.

\subsection{State Estimation and Observability}
\label{ss:state estimation}
The Kalman filter is designed for the setting where the underlying dynamical system is linear, and disturbances are Gaussian. In particular, consider a linear-time-invariant (LTI) autonomous system with state and measurement disturbances: let $x_t \in \R^n$ be the system state, $w_t \in \R^n$ the process noise, $y_t \in \R^p$ the measurement, and $v_t \in \R^p$ the measurement noise. The initial condition, process noise, and measurement noise are assumed to be i.i.d.\ zero-mean Gaussians: $x_0 \sim{} \calN(0,\Sigma_0)$, $w_t \overset{\mathrm{i.i.d.}}{\sim} \calN(0,\Sigma_w)$, $v_t \overset{\mathrm{i.i.d.}}{\sim} \calN(0,\Sigma_v)$. The LTI system is then defined by:
\begin{equation}\label{eq:ss}
    \begin{aligned}
     x_{t+1} &= Ax_t + w_t, \\
     y_t &= C x_t + v_t.
\end{aligned}
\end{equation}
Finite horizon state estimation determines an estimate for the state of the system at time $k$ given some sequence of measurements $y_0, \dots, y_N$.  This problem encompasses smoothing ($k<N$), filtering ($k=N$), and prediction ($k>N$). When the measurement and process noise satisfy the assumptions above, the optimal state estimator is the celebrated Kalman filter (or smoother/predictor), which produces state estimates that are a linear function of the observations.  Therefore, the optimal estimate $\hat{x}_k$ of the state $x_k$ at time $k$ can be written as\footnote{State-space representations for the Kalman filter (see \Cref{ss: kalman state space}) also exist \citep{hassibi1999}, but for our purposes it is more convenient to view it as a linear map.} $\hat{x}_k:= L Y_N$, where $L\in \R^{n\times p(N+1)}$ is some matrix and $Y_N$ is a vector of stacked observations 
\[
    Y_N := \bmat{y_0 & \hdots & y_N }^\top.
\]
We similarly define the stacked process and measurement noise vectors as
\[
    W_N := \begin{bmatrix} w_0 & \hdots & w_{N-1} \end{bmatrix}^\top \mbox{ and }  V_N := \bmat{v_0 & \hdots & v_N}^\top.
\]
Furthermore, suppose $k \leq N$ and let
\begin{align*}
    \calO_N &= \bmat{ C \\ CA \\ \vdots \\ CA^N}, \quad \tau_N = \bmat{0 & & & & \\ C & 0  & & \\  CA & C &  && \\ \vdots  & & \ddots & 0 & \\ CA^{N-1} & \ldots & & C & 0 } \\
    \Gamma_k &= \bmat{A^{k-1} & A^{k-2} & \ldots & I & 0 & \ldots & 0} 
\end{align*}
so that $Y_N=\calO_N x_0 + \tau_N W_N + V_N$ and $x_k = A^k x_0 + \Gamma_k W_N$. Here $\calO_N$ is the $N$-step observability matrix, which is a quantity of interest in our analysis. Recall that a system of the form \eqref{eq:ss} is observable if and only if the observability matrix $\calO_{n-1}$ has rank $n$.

Throughout the remainder of the paper we make the following assumption:
\begin{assumption}
    \label{as: observable}
    System \eqref{eq:ss} is observable and $N \geq n-1$. 
\end{assumption}

As stated, observability is a binary notion that determines whether state consistent estimation is possible. However, observability does not capture the conditioning of the problem defining the Kalman filter.

A more refined, non-binary notion of observability can be defined in terms of the \emph{observability gramian} of a system.
\begin{definition}
    The \emph{$N$-step observability gramian} is defined as $W_o(N) := \calO_N^\top \calO_N$. If the spectral radius of $A$ is less than one, then taking the limit as $N \to \infty$ results in the \emph{observability gramian} $W_o(\infty) = \sum_{t=0}^\infty \paren{A^t}^\top C^\top C A^t$. 
\end{definition}

The observability gramian provides significantly more information about the difficulty of state estimation than the rank condition on the observability matrix. In particular, the ellipsoid $\curly{x \, \vert \, x^\top W_o(\infty)x \leq 1}$ contains the initial states $x$ that lead to measurement signals with $\ell^2$ norm bounded by $1$ in the absense of process and measurement noise. To see this, let $x_0 = x$. Then we have $x_0^\top W_o(\infty)x_0 = \sum_{t=0}^\infty x_0^\top \paren{A^t}^\top C^\top C A^t x_0 = \sum_{t=0}^\infty x_t^\top C^\top C x_t = \sum_{t=0}^\infty \norm{y_t}_2^2$. As such, small eigenvalues of the observability gramian imply that a large subset of the state space leads to relatively small impacts on future measurements. This makes it difficult to use measurements to  distinguish states in this region in the presence of process and measurement noise, requiring high-gain estimators. This in turn suggests that such estimators may be more susceptible to small adversarial perturbations.

\subsection{Kalman Filtering and Smoothing}
\label{ss:kalman}

We begin by reviewing relevant results from standard Kalman Filtering and Smoothing.  
\paragraph{Standard State Estimation} Under Assumption \ref{as: observable}, we define the minimum mean square estimator for the state $x_k$ as
\begin{align}\label{eq:sr-kf}
    \hatL_k = \argmin_{L \in \R^{n\times p(N+1)}} \Ex \brac{\norm{x_k - L Y_N}_2^2}.
\end{align}
We note that the optimal solution to this problem is precisely the Kalman filter ($k=N$) or smoother ($k<N$). 
 We explicitly solve for the minimum mean square estimator $\hatL_k$ in the following standard lemma, included for completeness. 

\begin{lemma}
    \label{lem:kf}
    Suppose $k \leq N$. The finite horizon Kalman state estimator is the solution to optimization problem \eqref{eq:sr-kf}, and is given by 
    \begin{align*}
        \hatL_k &=
        \paren{ A^k \Sigma_0 \calO_N^\top +  \Gamma_k \Sigma_w \tau_N^\top}\cdot\paren{\calO_N \Sigma_0 \calO_N^\top +
        \tau_N \Sigma_w \tau_N^\top + \Sigma_v }\I.
    \end{align*}
\end{lemma}

\paragraph{Adversarially Robust State Estimation}
We now modify the standard filtering problem \eqref{eq:sr-kf} to allow adversarial perturbations to enter through sensor measurements.\footnote{We choose to restrict our attention to adversarial sensor measurements because it is a more direct analog to the traditional adversarial robustness literature, which considers perturbations to image data, and not to the image data-generating distribution \citep{szegedy2013intriguing, goodfellow2014explaining, carlini2016defensive}.} In particular, for some $\varepsilon>0$, the adversarially robust state estimator is defined by
\begin{align}
    \label{eq: advRobust Filter}
    \hatL_k(\varepsilon):= \argmin_{L \in \R^{n\times p(N+1)}} \Ex \brac{\max_{\norm{\delta}_2 \leq \varepsilon} \norm{x_k - L (Y_N + \delta)}_2^2}. \hspace{-0.2cm}
\end{align}


In contrast to the nominal state estimation problem, no closed form expression exists for the adversarially robust estimation problem, due to the inner maximization in \eqref{eq: advRobust Filter}. We show next that despite the non-convexity of the inner maximization problem, it can be solved efficiently.  This allows us to apply stochastic gradient descent to solve  for $\hat L_k(\varepsilon)$. In particular, note that the objective to the minimization problem is the expectation of a point-wise supremum of convex functions in $L$, and hence convex in $L$ itself \citep{boyd2004convex}. Next observe that we can draw samples of $x_0 \sim \calN(0,\Sigma_0)$, $W_N\sim \calN(0,I_N \otimes \Sigma_w)$, $V_N \sim \calN(0, I_{N+1} \otimes \Sigma_v)$, and apply the solution to the inner maximization problem to solve for realizations of $\max_{\norm{\delta}_2 \leq \varepsilon} \norm{x_k - L(Y_N+\delta)}_2^2$. Taking the gradient of these realizations with respect to $L$ provides a stochastic descent direction. As the overall expression is convex in $L$, stochastic gradient descent with an appropriately decaying stepsize converges to the optimal solution \citep{bottou2018first}.

\subsection{Solving the Inner Maximization}
\label{sec: inner max}

As earlier stated, no closed-form expression exists for the adversarially robust estimation problem: indeed, even in scalar linear regression studied in \citep{javanmard2020precise}, it is characterized by a recursive relationship. Furthermore, the techniques used to derive that recursion do not extend to the multi-variable case.

To address this challenge, we show how to efficiently compute solutions to the inner maximization in \eqref{eq: advRobust Filter}.  We observe that the maximization $\max_{\norm{\delta}_2 \leq \varepsilon} \norm{x_k - L(Y_N + \delta)}_2^2$ can be expanded and re-written as the following (non-convex) quadratically-constrained quadratic maximization problem:
\begin{align*}\label{eq:P}
    \maximize_{\delta \in \R^n}&\quad \delta^\top L^\top L \delta - 2\delta^\top L^\top b \tag{P}\\
    \subjectto &\quad \delta^\top \delta \leq \varepsilon^2,
\end{align*}
where we set $b := x_k - LY_N$.  Let $L = U \bmat{\Sigma & 0} V^\top \in \R^{n \times (N+1)p}$ be the full singular-value decomposition of $L$, with $U \in \R^{n\times n}$, $\Sigma \in \R^{n \times (N+1)p}$, $V\in \R^{(N+1)p \times (N+1)p}$, and $\Sigma = \mathrm{diag}(\sigma_1,\dots,\sigma_n)$, $\sigma_1 \geq \cdots \geq \sigma_n \geq 0$ the singular values of $L$. We also denote the columns of $U$ and $V$ by $u_i$ and $v_i$, respectively. It is known that \eqref{eq:P} satisfies strong duality \citep{boyd2004convex} and the optimal primal-dual pair $(\delta^*,\lambda^*)$ can be characterized by the KKT conditions:
\begin{align*}
    2(\lambda^* I-L^\top L)\delta^* + 2L^\top b &= 0 \\
    \lambda^* ({\delta^*}^\top \delta^* - \varepsilon^2) &= 0 \\
    (\lambda^* I -L^\top L) &\succeq 0.
\end{align*}
The KKT conditions can then be leveraged to solve for the optimal dual solution $\lambda^*$ and subsequently the optimal perturbation $\delta^*$. The full procedure is summarized in Algorithm~\ref{alg:getPerturbation}.  

We note that \citet{boyd2004convex} shows how to solve (P) via semidefinite programming. Algorithm~\ref{alg:getPerturbation}, however, allows us to recycle the SVD of $L$ to solve (P) for different values of $b := x_k - LY_N$ simply by solving a root finding problem for each $b$.  This enables efficient batching when applying SGD to the outer minimization problem. 

\begin{algorithm}
    \caption{Inner Maximization Solution}
    \label{alg:getPerturbation}
    \begin{algorithmic}
        \State \textbf{given} $L=U\Sigma V^\top \in \R^{n\times (N+1)p}$, $b\in\R^n$, perturbation bound $\varepsilon > 0$
        \If {$\sum_{i:\sigma_i < \sigma_1} \frac{(b^\top u_i)^2 \sigma_i^2}{(\sigma_1^2 - \sigma_i^2)^2} < \varepsilon^2$}
            \State $c = \sqrt{\varepsilon^2 - \sum_{i:\sigma_i < \sigma_1} \frac{(b^\top u_i)^2 \sigma_i^2}{(\sigma_1^2 - \sigma_i^2)^2}}$
            \State Set $v$ as any unit vector lying in the null-space of $\paren{\sigma_1^2 I -\Sigma^\top\Sigma }V^\top$, i.e.\ $v\in\Span\curly{v_i: \sigma_i = \sigma_1}$
            \State $\delta^* = -V (\sigma_1^2 I-\Sigma^\top\Sigma)^{\dagger} \Sigma^\top U^\top b + cv$
        \Else 
            \State solve $\sum_{i=1}^n \frac{(b^\top u_i)^2 \sigma_i^2}{(\lambda - \sigma_i^2)^2}=\varepsilon^2 $ for $\lambda$, e.g.\ by Newton's method
            \State $\delta^* = -V (\lambda^* I-\Sigma^\top\Sigma)^{-1} \Sigma^\top U^\top b$
        \EndIf
        \State \Return $\delta^*$
    \end{algorithmic}
\end{algorithm}

\noindent The proof of correctness for Algorithm \ref{alg:getPerturbation} is detailed in \Cref{ss:proof of inner max algorithm}. 


\section{Robustness-Accuracy Tradeoffs in Kalman Filtering}
\label{sec: Kalman Bounds}

The Kalman state estimation problem and adversarial state estimation problem can be viewed as standard and adversarially robust risk minimization problems by defining
\begin{align*}
    \SR(L) &:= \Ex \brac{\norm{x_k - L Y_N}_2^2}, \\
    \AR(L) &:= \Ex \brac{\max_{\norm{\delta}_2 \leq \varepsilon} \norm{x_k - L(Y_N + \delta)}_2^2}.
\end{align*}

Our goal is to characterize robustness-accuracy trade-offs for this linear inverse problem. We refer to the set of points $(\SR(L), \AR(L))$ over all $L \in \mathbb{R}^{n\times (N+1)p}$ as the $(\SR,\AR)$ region. The optimal tradeoff between
standard and adversarial risks is characterized via the so-called Pareto boundary of this region, which we denote $\curly{(\SR(L_\lambda),\AR(L_\lambda)): \lambda \geq 0}$. Using standard results in multi-objective optimization, $L_\lambda$ are computed by solving the regularized optimization problem
\begin{align}\label{eq:reg-opt}
    L_\lambda &:= \argmin_{L}\; \SR(L) + \lambda \AR(L).
\end{align}
Varying the regularization parameter $\lambda$ in problem \eqref{eq:reg-opt} thus allows us to characterize the aforementioned Pareto boundary by interpolating between the solution to the standard (i.e. $L_0$) and adversarial (i.e. $L_\infty$) problems. Via our results from Section~\ref{sec: inner max}, each solution $L_\lambda$ to the regularized optimization problem \eqref{eq:reg-opt} can be computed efficiently using stochastic optimization. We use these results to trace out the optimal tradeoff curves for specific examples in Section~\ref{sec: numerical}.

In this section, we show that the gap $\AR(L)-\SR(L)$ can be bounded in terms of the spectral properties of the observability gramian of the system, establishing a natural connection to the robust control and estimation literature \citep{hassibi1999,zhou1998robust}. In particular, our results indicate the robustness-accuracy tradeoff is more severe for systems with uniformly low observability, as characterized by the Frobenius norm of the observability gramian.
  


\subsection{Tradeoffs for Linear Inverse Problems}


We note that the Kalman state estimation problem can be posed as a general linear inverse problem, where 
$x \sim \calN(0, \Sigma_x)$, $w \sim \calN(0, \Sigma_w)$, $y = Mx+w$, and our goal is to minimize one of the following risks
\begin{align*}
    \SR(L) &= \Ex \brac{\norm{x - Ly}_2^2}, \\
    \AR(L) &= \Ex \brac{\max_{\norm{\delta}_2 \leq \varepsilon} \norm{x - L(y+\delta)}_2^2}.
\end{align*}

Although no closed-form expression exists for the adversarial risk $\AR(L)$ exists, we show now that interepretable upper and lower bounds on the robustness-accuracy tradeoff, as characterized by the gap $\AR(L) - \SR(L)$, can be derived.  Such bounds predict the severity of the robustness-accuracy tradeoff based upon underlying properties of specific linear inverse problems. We further show that these bounds are tight in the sense that they are exact for certain classes of matrices $L$, and strong in the sense that the lower and upper bounds differ only in higher-order terms with respect to the adversarial budget $\varepsilon$.

\begin{theorem}
\label{thm:generalUBLB}
Given any $L \in \R^{p \times n}$, we have the following lower bound on $\AR(L) - \SR(L)$:

\begin{equation}
\label{eq: s-lemma lower bound}
\begin{aligned}
    \AR(L) - \SR(L) &\geq 2\varepsilon\; \Ex_{x,w}\brac{\norm{L^\top (x - Ly)}_2} + \varepsilon^2 \lambda_{\min} (L^\top L),
\end{aligned}
\end{equation}
and a corresponding upper bound
\begin{equation}
\label{eq: s-lemma upper bound}
\begin{aligned}
    \AR(L) - \SR(L)  &\leq 2\varepsilon\; \Ex_{x,w}\brac{\norm{L^\top (x - Ly)}_2} + \varepsilon^2 \lambda_{\max}(L^\top L),
\end{aligned}
\end{equation}
where $\lambda_{\min}(L^\top L)$ and $\lambda_{\max}(L^\top L)$ are the minimum and maximum eigenvalues of $L^\top L$, respectively.
\end{theorem}

The proof of these bounds relies on turning the inner maximization of the adversarial risk into various equivalent optimization problems, and utilizing the properties of Schur complements and the S-lemma. See \Cref{ss:proof of general UBLB} for details.

We note that when $L^\top\in\R^{n}$, inequality \eqref{eq: s-lemma upper bound} recovers the exact characterization of the gap $\AR(L)-\SR(L)$ provided in \cite{javanmard2020precise}; thus when $p=1$, the inequality \eqref{eq: s-lemma upper bound} is in fact an equality.
We also note that the upper and lower bounds differ only in the  $\calO(\varepsilon^2)$ terms. This leads immediately to the following result.

\begin{corollary}
If $p \geq n$ and $L$ has orthogonal columns, then bounds \eqref{eq: s-lemma lower bound} and \eqref{eq: s-lemma upper bound} match.
\end{corollary}
\begin{proof}
If $p \geq n$ and $L$ has orthogonal columns, then $\lambda_{\min}(L^\top L)=\lambda_{\max}(L^\top L)$.
\end{proof}

\begin{figure}
    \centering
    \includegraphics[width=0.5\textwidth]{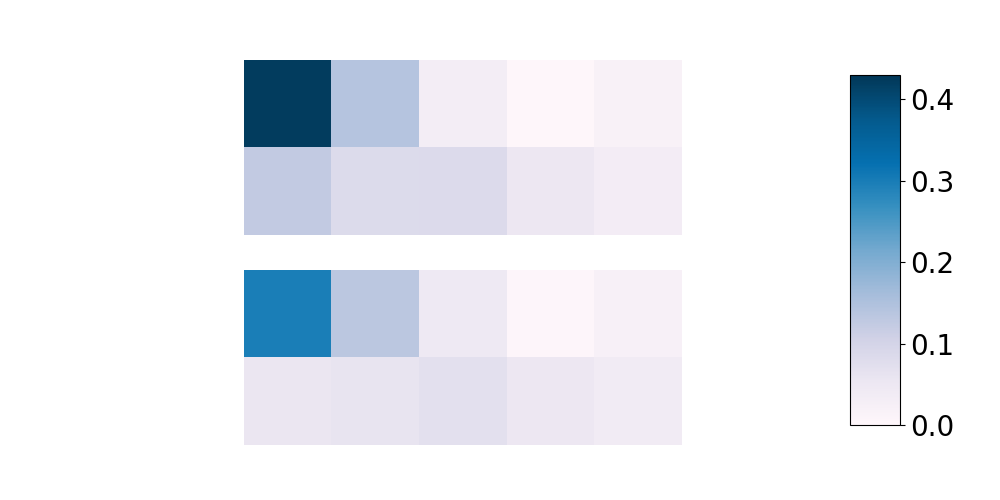}
    \vspace*{-0.5cm}
    \caption{Optimizing for adversarial robustness induces an implicit regularization, which is visualized in this heatmap of a $2\times5$ nominal initial state estimator $L_\star$ (top) and adversarially robust solution $\hat{L}(\varepsilon)$ (bottom), where $\varepsilon= 5$, $x_0 \sim \calN(0, I)$, $w_t \sim \calN(0, I)$, $v_t \sim \calN(0, 1)$, $A = \bmat{1 & 1 \\ 0 & 1}$, $C = \bmat{0, 1}$, and $N = 4$.}   
    
    \label{fig:heatmap}
\end{figure}

The terms involving the eigenvalues of $L^\top L$ in our bounds also support the intuition that adversarial robustness is a form of implicit regularization, which is visualized in Figure~\ref{fig:heatmap}. In the one-dimensional linear classification setting, this phenomenon is well-understood \citep{tsipras2018robustness, dobriban2020provable}, where robustness to adversarial perturbations prevent a robust feature vector from relying on an aggregate of small features. We note that in the case of state estimation, we have in general $p < n$, and thus the quadratic factor in $\varepsilon$ is $0$ in the lower bound~\eqref{eq: s-lemma lower bound}.


In the subsequent section, we will leverage bounds \eqref{eq: s-lemma lower bound} and \eqref{eq: s-lemma upper bound} from Theorem \ref{thm:AR-SR gap lower bound} to bound both the susceptibility and robustness of the Kalman Filter. 



\subsection{Bounding $\AR - \SR$ for State Estimation}

\label{ss:bounding state estimation error}
The adversarial risk does not admit a closed-form solution in general. Upper and lower bounds on the gap between the adversarial risk and standard risk, however, still highlight the role control theoretic quantities play in robustness-accuracy tradeoffs. We make the following simplifying assumption for presentation purposes going forward. 
\begin{assumption}\label{as:noisepd}
    $\Sigma_0 = \sigma_0^2 I $, $\Sigma_w = \sigma_w^2 I$, $\Sigma_v = \sigma_v^2 I$. We further assume that the system matrix $A = \rho Q$, $\rho \in [0,1]$, is a scaled orthogonal matrix, such that $\rho$ controls the stability of the system.
\end{assumption} 

As stated in \Cref{as: observable}, $(A,C)$ is always assumed to be observable. Generalizations of our subsequent results to generic dynamics $A$ and positive definite covariance matrices are stated and proven in \Cref{ss: kf general statements and proofs}: although more notationally cumbersome, they nevertheless convey the same overall trends.

We first present a closed form for the standard risk $\SR(L)$ which indicates the role that observability plays in robustness-accuracy tradeoffs. 
\begin{lemma} \label{lem:SR closed form}
The standard risk may be expressed as
    \begin{align*}
        \SR(&L) = \Ex \brac{\norm{x_k - L Y_N}_2^2} =\sigma_0^2\norm{A^k - L \calO_N}_F^2  + \sigma_w^2\norm{\Gamma_k - L \tau_N}_F^2 + \sigma_v^2\norm{L}_F^2.
    \end{align*}
\end{lemma}

\noindent Lemma \ref{lem:SR closed form} makes clear that the noise terms act as a regularizer: if $\sigma_w^2=\sigma_v^2=0$, then $\min_L \SR(L)=0$ and is achieved by $L=A^k(W_o(N))^{-1}\calO_N^\top$. The gain of this filter has clear dependence upon the spectral properties of $W_o(N)$, indicating the key role that $W_o(N)$ plays in the robustness-accuracy tradeoffs satisfied by an LTI system \eqref{eq:ss}. We formalize this intuition next.
As a first step, we specialize the lower bound in Theorem~\ref{thm:generalUBLB} to the dynamical system setting.  

\begin{lemma}\label{thm:AR-SR gap lower bound}
    For any $L \in \R^{n \times p(N+1)}$, the gap between $\AR(L)$ and $\SR(L)$ admits the following lower bound:
    \begin{align}\label{eq:AR-SR gap no exp}
         \AR(L) - \SR(L)
         \geq 2\sqrt{\frac{2 }{\pi}} \frac{\varepsilon}{\sqrt{n}} \sigma_v \norm{L}_F^2
    \end{align}

\end{lemma}

We now turn our attention to studying the tradeoffs enjoyed by the Kalman Filter/Smoother $L = \hat{L}_k$ defined in Lemma \ref{lem:kf}. Since the Kalman estimator is the optimal estimator in the nominal setting and is commonly used in practice, instantiating Lemma~\ref{thm:AR-SR gap lower bound} for $L = \hat{L}_k$ captures the susceptibility of a nominal estimator to small adversarial perturbations.
To simplify notation in the subsequent results, we will denote $\sigma_{\vee}^2 = \max\curly{\sigma_0^2, \sigma_w^2}$, $\sigma_{\wedge}^2 = \min\curly{\sigma_0^2, \sigma_w^2}$ and
    \begin{equation}
        r_k(\rho) = \begin{cases} k, \quad& \rho = 1 \\ \frac{1 - \rho^{2(k+1)}}{1 - \rho^2}, \quad & \rho \neq 1.
    \end{cases}
    \end{equation}
With these definitions, we have the following theorem. 

\begin{theorem}
     Suppose that $\hat{L}_k$ is the Kalman estimator from Lemma~\ref{lem:kf}. 
    We have the following bound on the gap between $\AR$ and $\SR$. 
    \begin{equation}
    \begin{aligned}\label{eq:lb}
        \AR(\hat{L}_k)& - \SR(\hat{L}_k) \geq 2\sqrt{\frac{2 }{\pi}} \frac{\varepsilon}{\sqrt{n}}\sigma_v \norm{C}_F^2 \paren{ \frac{\rho^{2k} \sigma_0^2 + r_k(\rho)\; \sigma_w^2 }{(N+1) \sigma_{\vee}^2\norm{W_o(N)}_F + \sigma_v^2 } }^2. 
    \end{aligned}
    \end{equation}
    \label{thm:lb}
\end{theorem}
We see that the lower bound increases as the Frobenius norm of the observability gramian decreases. This indicates that as observability becomes uniformly low, i.e., if all eigenvalues of $W_o(N)$ are small, then a nominal state estimator $\hat L_k$ will have a large gap $\AR(\hat L_k)-\SR(\hat L_k)$. Observe that increasing $\sigma_w$ will increase the lower bound shown above when $\sigma_w \leq \sigma_0$.

We now derive an upper bound on the gap between the standard and adversarial risk for any given $L$. This bound follows from the upper bound in Theorem \ref{thm:generalUBLB}.
\begin{lemma}\label{thm:general upper bound}
    For any $L \in \R^{n \times p(N+1)}$, the following bound holds
    \begin{align*}
        \AR(L) -  \SR(L) \leq 2\varepsilon  \norm{L}_2 \norm{\Sigma^{1/2}}_F + \varepsilon^2 \norm{L}_2^2,
    \end{align*}
    where $\Sigma^{1/2}$ is the symmetric square root of the covariance of $x_k - LY_N$. 
\end{lemma}

Again, we consider how this upper bound looks for the Kalman estimator $\hat{L}_k$. 

\begin{figure*}[t]
    \centering
    \makebox[\textwidth][c]{
        \subfloat[][Initial state estimation $k=0$]{
        \includegraphics[width = 0.5 \textwidth]{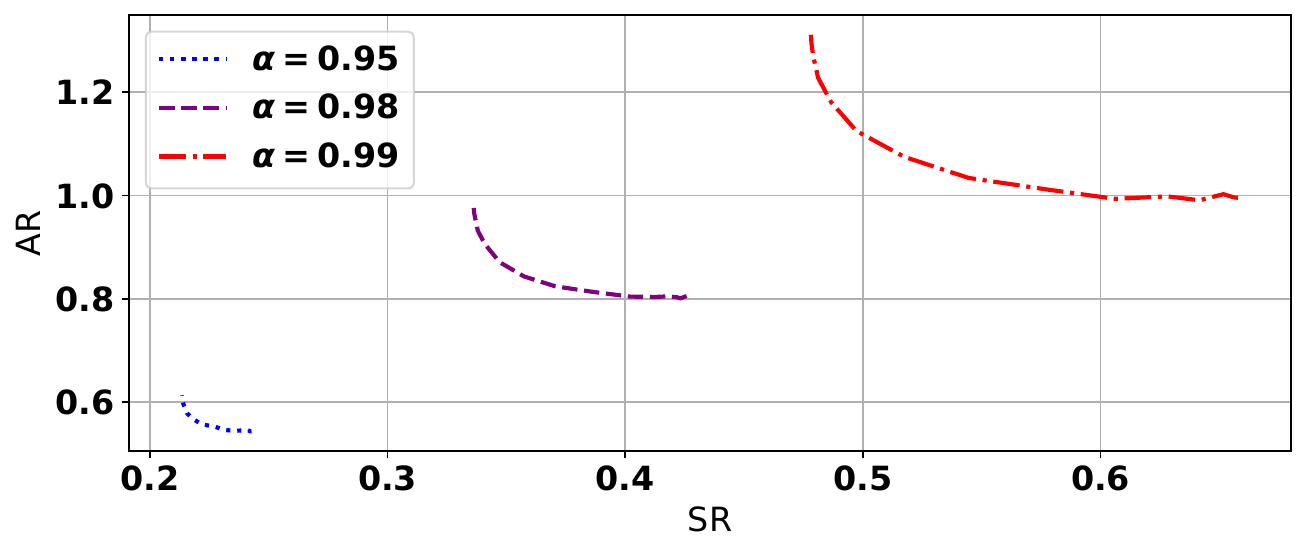}
        }
        \subfloat[][Final state estimation $k=N$]{
        \includegraphics[width = 0.5 \textwidth]{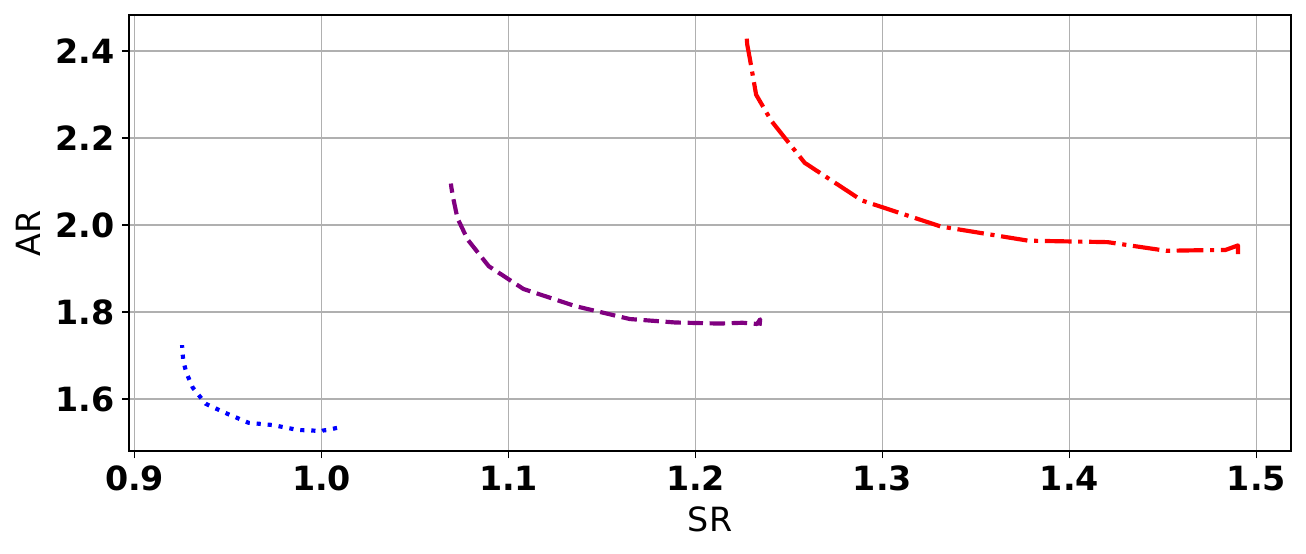}
        }
    }

    \caption{Pareto boundaries of $(\SR, \AR)$ for initial and final state estimation. The observability of a the system is determined by $\alpha$. When $\alpha$ approaches one, observability decreases and the tradeoff between $\SR$ and $\AR$ becomes more severe.}
    
    \label{fig:state estimation tradeoffs}
\end{figure*}

\begin{theorem}
    
    Suppose that $\hat{L}_k$ is the Kalman state estimator from Lemma \ref{lem:kf}. Then
    \begin{align*}
        \AR(\hat{L}_k)  - \SR(\hat{L}_k) \leq &\;\varepsilon \paren{\frac{ \rho^{2k} \sigma_0^2 + r_k(\rho)\; \sigma_w^2}{\sigma_{\wedge}^2 \lambda_{\min}(W_o(N))^{1/2}}}  \\
        &\cdot \Bigg[2 \sqrt{n}
            \paren{\sigma_{\vee}^2 + \paren{\frac{\sigma_v}{\sigma_{\wedge}^2\lambda_{\min}(W_o(N))^{1/2} }}^2 }^{1/2} +\varepsilon \paren{ \frac{1}{\sigma_{\wedge}^2 \lambda_{\min}(W_o(N))^{1/2} }}\Bigg].
    \end{align*}
    
    Furthermore, when $\lambda_{\min}(W_o(N))^{1/2} \geq \frac{\sigma_v}{\sigma_{\wedge}^2}$, we have
    \begin{align*}
        \AR(\hat{L}_k) - \SR(\hat{L}_k) 
        \leq &\;\varepsilon \paren{\frac{\sigma_{\wedge}^2 \lambda_{\min}(W_o(N))^{1/2}}{\sigma_{\wedge}^4 \lambda_{\min}(W_o(N)) + \sigma_v^2 }\paren{ \rho^{2k} \sigma_0^2 + r_k(\rho)\; \sigma_w^2} } \\
        & \cdot \Bigg[2 \sqrt{n}
            \paren{\sigma_{\vee}^2 + \sigma_v^2 \paren{\frac{\sigma_{\wedge}^2 \lambda_{\min}(W_o(N))^{1/2}}{\sigma_{\wedge}^4 \lambda_{\min}(W_o(N)) + \sigma_v^2 }}^2 }^{1/2} +\varepsilon \paren{\frac{\sigma_{\wedge}^2 \lambda_{\min}(W_o(N))^{1/2}}{\sigma_{\wedge}^4 \lambda_{\min}(W_o(N)) + \sigma_v^2 }}\Bigg].
    \end{align*}
    
    \label{thm:ub}
\end{theorem}
The upper bound on the gap decreases as the minimum eigenvalue of the observability gramian increases. This indicates that as the observability of the system becomes uniformly large, the gap between standard and adversarial risk for the nominal Kalman estimator will decrease. Perhaps counter-intuitively, when observability is poor in some direction, i.e. when $\lambda_{\min}\paren{W_o(N)}$ is small, increasing the sensor noise $\sigma_v$ will actually \emph{decrease} the above upper bound, as long as $\lambda_{\min}\paren{W_o(N)} \geq \frac{\sigma_v}{\sigma_\wedge^2}$. This aligns with results demonstrating that injecting artificial noise can improve the robustness of state observers \citep{doyle1979robustness}, and is further consistent with our interpretation of noise as a regularizer following Lemma \ref{lem:SR closed form}.

We note that since the properties of the observability gramian $W_o(N)$ are tied to $\rho$, it is not immediately clear how to extract the role of stability $\rho \in [0,1]$ in either \Cref{thm:lb} or \Cref{thm:ub}. This is to be expected, as the fragility of the Kalman Filter has more to do with the observability of the system rather than its autonomous stability. For example, when $C^\top C = I$, corresponding to maximal observability $W_o(N) = r_k(\rho)$, and $k \leq \calO(N)$, then the dominant terms in the lower and upper bound are essentially independent of $\rho$ when $N$ is large.

\section{Numerical Results}
\label{sec: numerical}

We now demonstrate that the theoretical results shown in the previous section predict the tradeoffs arising in Kalman Filtering problems. 


\begin{figure}
    \centering
    \includegraphics[width=0.5\textwidth]{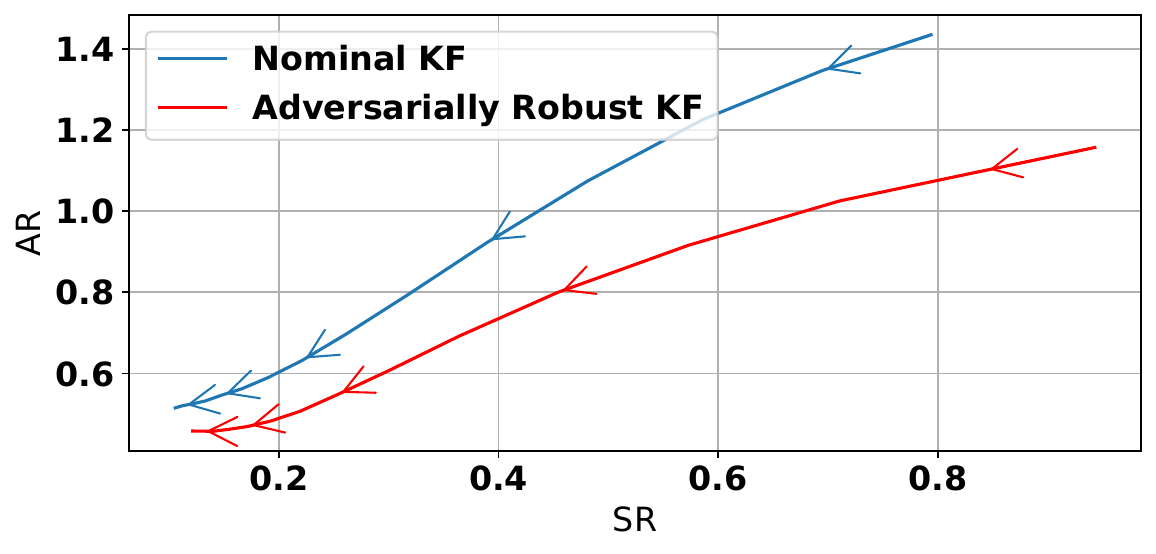}
    \vspace{-12pt}
    \caption{$\SR$ vs $\AR$ for a nominal Kalman smoother and an adversarially robust smoother, where observability of the underlying system increases in the direction of the arrows. For any fixed value of the standard risk, the adversarially robust smoother achieves a  lower adversarial risk than the nominal smoother. On the right side of the plot, when observability is low, the adversarial risk of the adversarially robust smoother is much lower than that of the nomimal smoother. This difference shrinks as we move to the left and observability increases.} 
    \label{fig:kf vs adv kf}
\end{figure}


\paragraph{Pareto Curves for Adversarial Kalman Filtering} In this experiment we compute the Pareto-optimal frontier for adversarially robust  Kalman filtering on systems with varying observability. We consider the system defined by the tuple
$\paren{A, C, \Sigma_0, \Sigma_w, \Sigma_v, N} = \paren{\bmat{\alpha & \beta \\ -\beta & \alpha}, \bmat{1 & 0}, I, 0.1 I, 0.1, 5}$ where $\alpha^2 + \beta^2 =1$, and we vary $\alpha$. As $\alpha$ approaches one the minimum eigenvalues of the observability gramian become small. In particular,  for $\alpha = 0.95$, $\alpha=0.98$, $\alpha=0.99$, the minimum eigenvalues of the observability gramian are given by $1.22, 0.81$ and $0.58$ respectively. The adversarial budget is set to $\varepsilon = 0.5$. Figure~\ref{fig:state estimation tradeoffs} shows the resulting tradeoff curves which demonstrate that as observability decreases, both $\SR$ and $\AR$ increase, as do the distance between the extremes of the Pareto curve. 
The results therefore support Section~\ref{ss:bounding state estimation error}, where we showed shrinking the eigenvalues of $W_o(N)$ increases the severity of the tradeoff between $\SR$ and $\AR$.  

 \begin{figure*}[t]
    \centering
    \makebox[\textwidth][c]{
        \subfloat[][Linear estimator]{
        \includegraphics[width = 0.5 \textwidth]{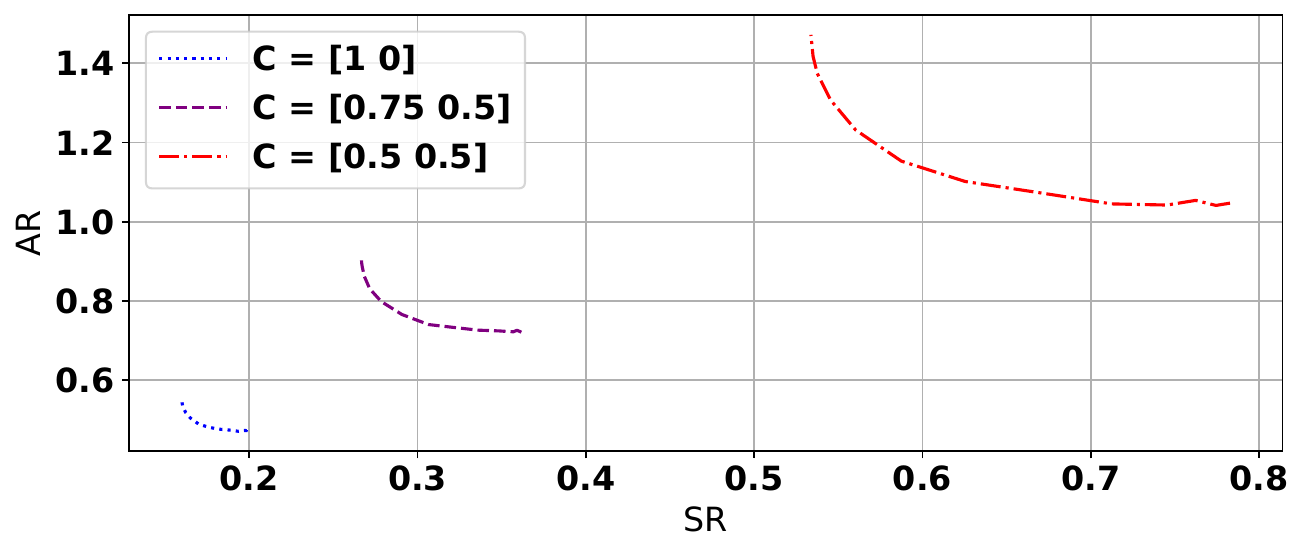}
        }
        \subfloat[][Neural Network]{
        \includegraphics[width = 0.5 \textwidth]{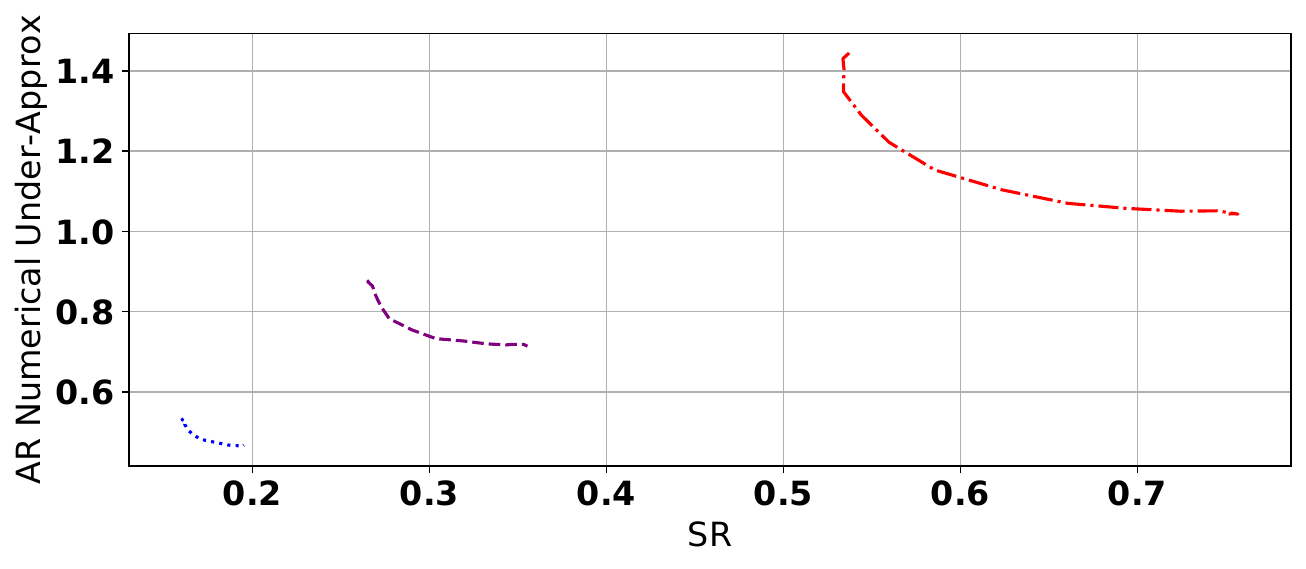}
        }
    }
    \caption{Pareto boundaries of $(\SR, \AR)$ for initial state estimation for a variety of measurement matrices $C$ by both a linear state estimator and a neural network (Lower bound on $\AR$ plotted for NN). As the first entry of $C$ decreases, the tradeoff curve becomes more severe. The trade-offs are not alleviated by a nonlinear estimator.} 
    \label{fig:NN tradeoffs}
\end{figure*}

\paragraph{Tradeoffs of Kalman Smoother versus Adversarially Robust Kalman Smoother} In Figure \ref{fig:kf vs adv kf}, we demonstrate the impact of adversaries on the risk incurred by an estimator optimized for $\SR$ versus $\AR$.
 We consider initial state estimation of a system defined by the tuple $(A, C, \Sigma_0, \Sigma_w, \Sigma_v, N) = \paren{\bmat{1 & \rho \\ 0 & 1}, \bmat{1 & 0}, I, 0.1 I, 0.1, 5}$, where we vary $\rho$ from $0.1$ to $\sqrt{10}$, such that the eigenvalues of $W_o(N)$ increase as $\rho$ increases. The adversarial budget is fixed at $\varepsilon = 0.5$.  Evaluating the nominal and robust estimators on this class of systems, we see that the adversarially robust smoother has significantly smaller adversarial risk compared to the nominal Kalman smoother when observability is low. As observability increases, this advantage shrinks. This suggests that considering the tradeoffs from adversarially robust state estimation is most important when observability is low. 
  \begin{figure*}[t]
     \centering
     \makebox[\textwidth][c]{
        \subfloat[][Nominal Setting]{
        \includegraphics[width = 0.48 \textwidth]{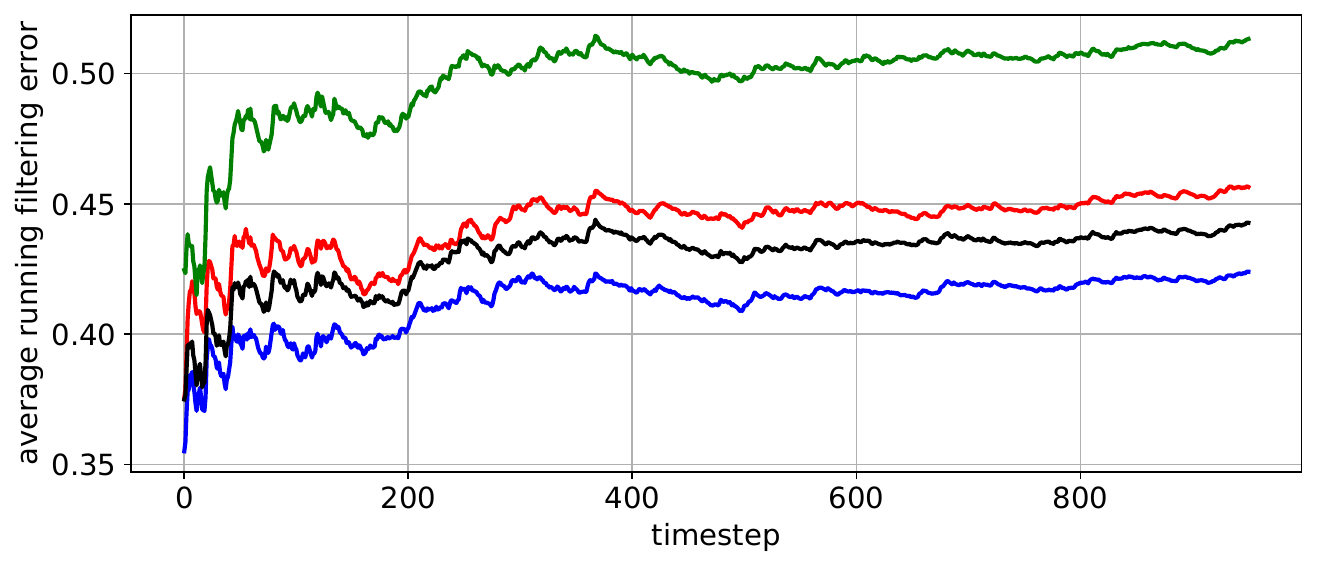}
        }
        \subfloat[][Sensor Drift]{
        \includegraphics[width = 0.48 \textwidth]{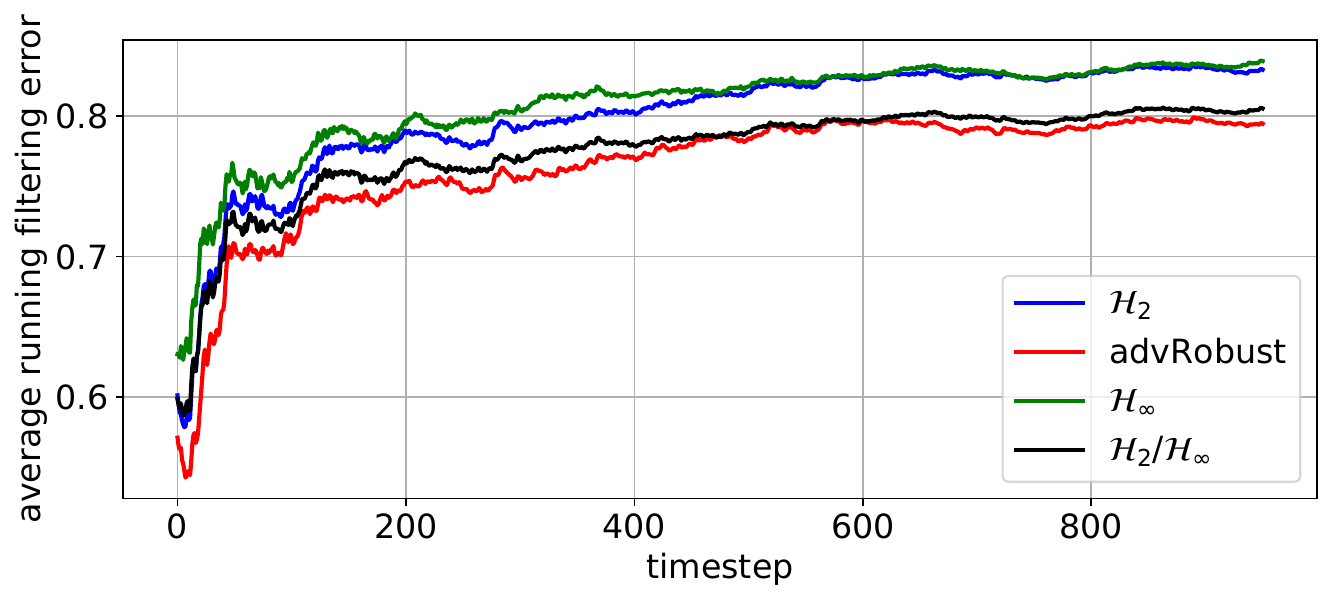}
        }
    }
     \caption{Simulations of the performance of the Kalman/$\mathcal{H}_2$ filter, our adversarially robust Kalman filter, a mixed $\mathcal{H}_2/\mathcal{H}_\infty$ filter, and the $\mathcal{H}_\infty$ filter on a linear system. The average filtering error is computed as $\frac{1}{t} \sum_{k=1}^t \norm{x_t - \hat x_t}_2^2$.}
     \label{fig:advkf performance}
 \end{figure*}

 \paragraph{Tradeoffs with Nonlinear State Estimators}
 In Figure~\ref{fig:NN tradeoffs}, we demonstrate empirically that the fundamental tradeoffs are not overcome by using nonlinear state estimators. In particular, we consider the system defined by  $(A, \Sigma_0, \Sigma_w, \Sigma_v, N) = \paren{\bmat{1 & 1 \\ 0 & 1}, I, 0.1 I, 0.1, 5}$, and $C$ as in the legend of the plots. We set the adversarial budget to $\varepsilon = 0.5$. We solve for the Pareto boundary for the linear estimator as in Figure~\ref{fig:state estimation tradeoffs}. We also consider a two layer network with ten neurons per layer and ReLU activation functions. To train this network, we perform an SGD procedure similar to that used in the linear case, with the exception being that we do not solve the exact adversary corresponding to each data point, but rather apply 100 steps of projected gradient ascent to find the adversarial perturbation. Once the neural network is trained, this same approach to find the adversary is used again to estimate the adversarial risk of the resultant estimator. As a result, the tradeoff curves shown for the neural network are an under-approximation to the true values of $\AR$. As such, it appears that the neural network is getting roughly the same tradeoff curves as the linear estimator, which is expected due to the fact that linear estimators are optimal in the $\mathcal{H}_2$ and $\mathcal{H}_\infty$ settings \citep{zhou1998robust}.

\paragraph{Comparison with Kalman Filter, mixed $\mathcal{H}_2/\mathcal{H}_\infty$, and $\mathcal{H}_\infty$ Filters}
 In Figure~\ref{fig:advkf performance}, we compare the performance of the adversarially robust Kalman filter with the optimal $\mathcal{H}_\infty$ filter, a mixed $\mathcal{H}_2/\mathcal{H}_\infty$ filter which minimizes the $\mathcal{H}_2$ norm subject to a $\mathcal{H}_\infty$ norm bound of $1.7$, and the nominal Kalman or $\mathcal{H}_2$ filter. In particular, we consider the filtering problem for a trajectory generated by the system $(A,C, \Sigma_0, \Sigma_w, \Sigma_v) =  \paren{\bmat{0.9 & 1 \\ 0 & 0.9}, \bmat{1, 0}, I, 0.1 I, 0.1}$. For ease of computation, we use the stationary mixed, $\mathcal{H}_\infty$, and Kalman filters \citep{lewis2008optimal, caverly2019lmi}, and the adversarially robust Kalman Filter for a horizon of 10 with adversarial budget $\varepsilon =0.75$.  We consider a nominal setting, where process and state disturbances are Gaussian with covariances defined by $\Sigma_w$ and $\Sigma_v$, and a setting to simulate sensor drift, where we have the white noise disturbances in addition to a sinusoidally varying measurement disturbance $\sin\paren{\frac{2 \pi k}{10}}$. The $\mathcal{H}_\infty$ norm bound for the mixed filter was tuned to the value of $1.7$ for good performance on the sensor drift setting. As expected, the nominal Kalman filter performs best in the nominal setting with zero-mean disturbances. In the setting with sensor drift, however, the adversarially robust Kalman filter performs substantially better than the nominal and $\calH_\infty$ filters, and slightly better than the mixed $\mathcal{H}_2/\mathcal{H}_\infty$ controller. In both settings, the $\mathcal{H}_\infty$ filter is overly conservative. Note that a key advantage of the adversarially robust controller is the interpretability of the robustness level determined by the parameter $\varepsilon$, which directly corresponds to the power of the adversarial disturbance. This feature is not shared by other filter designs, such as the $\mathcal{H}_\infty$ norm bound in the mixed $\mathcal{H}_2/\mathcal{H}_\infty$ filter. 
\section{Conclusion}
\label{sec:conclusion}

We analyzed the robustness-accuracy tradeoffs arising in Kalman Filtering. We did this in two parts. We first provided an algorithm to solve for the optimal adversarial perturbation, which can be used to trace out the Pareto boundary. We then bounded the gap between the adversarial and standard state estimation error in terms of the spectral properties of the observability gramian. These bounds extend upon the robustness-accuracy tradeoff results arising in the classification and linear regression settings. An interesting avenue of future work is to extend these results to infinite horizon filtering, and combine them with tradeoff analyses in LQR control \citep{lee2022performance} to provide an understanding of the adversarial tradeoffs in the control of partially observed linear systems. It would also be interesting to consider the adversarial tradeoffs of state estimation when adversarial state perturbations are also present. 

\section*{Acknowledgements}
Bruce Lee is supported by the Department of Defense through the National Defense Science \& Engineering Graduate Fellowship Program. The research of Hamed Hassani is supported by NSF Grants 1837253, 1943064, 1934876, AFOSR Grant FA9550-20-1-0111, and DCIST-CRA. Nikolai Matni is funded by NSF awards CPS-2038873, CAREER award ECCS-2045834, and a Google Research Scholar award. 

\bibliographystyle{abbrvnat}
\bibliography{refs}

\clearpage
\appendix

\section{Kalman Filtering State Space Solution} \label{ss: kalman state space}
Consider the setting defined in \Cref{ss:state estimation}, i.e.\ we have a dynamical system which progresses according to 
\begin{align*}
    x_{t+1} &= A x_t + w_t \\
    y_t &= C x_t + v_t
\end{align*}
with $x_0 \sim{} \calN(0,\Sigma_0)$, $w_t \overset{\mathrm{i.i.d.}}{\sim} \calN(0,\Sigma_w)$, $v_t \overset{\mathrm{i.i.d.}}{\sim} \calN(0,\Sigma_v)$. 
Consider estimating state $x_k$ given measurements $y_0, \dots, y_k$. The minimum mean square estimator is given by
\[
    \hat x_k = \min_z \Ex \brac{\norm{z - x_k}_2^2 | y_0, \dots, y_k}.
\]
This can be written as an integral
\[
 \min_z \int_{\R^n} \norm{z - x_k}_2^2 f(x_k | y_0, \dots, y_k) dx_k,
\]
where $f$ denotes the conditional density of $x_k$ given the measurements. The objective is convex in $z$, and thus we can find the minimizer by setting the gradient with respect to $z$ to zero. In particular, we have
\begin{align*}
    \frac{d}{dz} \int_{\R^n} \norm{z - x_k}_2^2 f(x_k | y_0, \dots, y_k) &=  \int_{\R^n} \frac{d}{dz} \norm{z - x_k}_2^2 f(x_k | y_0, \dots, y_k) dx_k\\
    &=\int_{\R^n} 2 (z - x_k) f(x_k | y_0, \dots, y_k) dx_k \\
    &=0,
\end{align*}
where dominated convergence theorem permits the exchange of integration and differentiation. Therefore, the state estimate may be expressed
\[
    \hat x_k = \int_{\R^n} x_k f(x_k | y_0, \dots, y_k) dx_k = \Ex\brac{x_k | y_0, \dots y_k}.
\]
Thus we can determine the state estimates $\hat x_k$ as the mean of the conditional distribution $f(x_k|y_0, \dots y_k)$. This may be computed recursively. In particular, let $x_k|y_{0:k}$ be the random variable with probability density function $f(x_k|y_0, \dots y_k)$. Then for all $k$, we have that $x_k|y_{0:k}= \calN(\hat x_k^+, P_k^+)$ where 
\begin{equation}
\begin{aligned}
\label{eq:KFequations}
    x_0^- &= 0\\ 
    P_0^- &= \Sigma_0 \\ 
    \hat x_k^+ &= x_k^- + P_k^- C^\top \paren{C P_k^- C^\top + \Sigma_v}^{-1} \paren{y_k - C\hat x_k^-} \\
    P_{k}^+ & = P_{k}^- - P_k^- C^\top \paren{C P_k^- C^\top + \Sigma_v} C P_k^- \\
    \hat x_{k+1}^- &= A \hat x_k^+\\
    P_{k+1}^- &= A^\top  P_k^+ A + \Sigma_w. \\
\end{aligned}
\end{equation}
To see that this is the case, recall that $x_0 \sim \calN(0, \Sigma_0)$, by assumption. Now suppose that $x_k|y_{0:k-1} \sim \calN(\hat x_k^-, P_k^-)$. Observe that 
\[
    \bmat{ x_k|y_{0:k-1} \\ y_k} \sim \calN\paren{ \bmat{\hat x_k^- \\ C \hat x_k^-}, \bmat{P_k^- & P_k^- C^\top \\ C P_k^- & C P_k^- C^\top + \Sigma_v}}
\]
Thus
\begin{align*}
    x_k|y_{0:k} &\sim \calN\paren{x_k^- + P_k^- C^\top \paren{C P_k^- C^\top + \Sigma_v}^{-1} \paren{y_k - C\hat x_k^-} ,P_{k}^- - P_k^- C^\top \paren{C P_k^- C^\top + \Sigma_v} C P_k^-} \\
    &= \calN(x_k^+, P_k^+)
\end{align*}
Now, given $x_k| y_{0:k} \sim \calN(x_k^+, P_k^+)$, observe that $x_{k+1}|y_{0:k} = A x_k|y_{0:k} + w_k \sim \calN(A^\top x_k^+,  A^\top P_k^+ A + \Sigma_w) = \calN(x_{k+1}^-, P_k^-)$.

Using the equations in \eqref{eq:KFequations}, we can write the Kalman filter as a state space system with inputs $y_t$. In particular, if $P_k^-$ and $P_k^+$ are defined as in \eqref{eq:KFequations}, we can let $K_k = P_k^- C^\top \paren{C P_k^- C^\top + \Sigma_v}^{-1}$. Then we may express our state estimates using the following time varying system.
\begin{align*}
    \hat x_{k+1} &= (A - K_k C A) \hat x_k + K_k y_k.
\end{align*}
\section{Proofs from Section \ref{sec: Kalman}}

\subsection{Proof of Lemma \ref{lem:kf}}
\textbf{Lemma~\ref{lem:kf}:}\;\:Suppose $k \leq N$. The finite horizon Kalman state estimator is the solution to optimization problem \eqref{eq:sr-kf}, and is given by 
    \begin{align*}
        \hatL_k &= \paren{A^k \Sigma_0 \calO_N\T + \Gamma_k \Sigma_w \tau_N\T}  \paren{\calO_N \Sigma_0 \calO_N\T+\tau_N\paren{I_N \otimes \Sigma_w} \tau_N\T +\paren{I_{N+1} \otimes \Sigma_v}}^{-1}.
    \end{align*}
\textit{Proof:} We know $\SR(L)$ is convex in $L$, thus we may take the derivative of $\SR(L)$ with respect to $L$ and set it to $0$ to solve for $\hatL_k$. Matrix derivatives can be found in \citet{petersen2008matrix}.

\subsection{Proof of Correctness for Algorithm \ref{alg:getPerturbation}} \label{ss:proof of inner max algorithm}

Recalling the optimization problem 
\begin{align*}
    \maximize_{\delta \in \R^n}&\quad \delta^\top L^\top L \delta - 2\delta^\top L^\top b \tag{P}\\
    \subjectto &\quad \delta^\top \delta \leq \varepsilon^2,
\end{align*}
and the corresponding KKT conditions:
\begin{align*}
    2(\lambda^* I-L^\top L)\delta^* + 2L^\top b &= 0 \\
    \lambda^* ({\delta^*}^\top \delta^* - \varepsilon^2) &= 0 \\
    (\lambda^* I -L^\top L) &\succeq 0.
\end{align*}
The third condition implies that $\lambda^* \geq \sigma_1^2$. We assume $\sigma_1 > 0$, otherwise the problem is trivial. Then using the SVD of $L$ to re-arrange the first stationarity condition, we get
\begin{align*}
    (\lambda^* I-\Sigma^\top \Sigma) V^\top \delta^* &= \Sigma^\top U^\top b.
\end{align*}
Maximizing a convex function over a convex set achieves its maximum on the boundary; it suffices to search over $\delta^\top \delta = \varepsilon^2$. We now consider two cases: $\lambda^* > \sigma_1^2$ and $\lambda^* = \sigma_1^2$. 

\begin{itemize}
    \item In the first case $\lambda^* > \sigma_1^2$, we know $\lambda^* I-\Sigma^\top \Sigma$ must be invertible, and thus
    \begin{align*}
        \delta^* &= -V (\lambda^* I-\Sigma^\top \Sigma)^{-1} \Sigma^\top U^\top b \\ 
        \varepsilon^2 = {\delta^*}^\top \delta^* &=  b^\top U \Sigma (\lambda^* I - \Sigma^\top \Sigma)^{-2} \Sigma^\top U^\top b \\
        &= \sum_{i=1}^{\min\curly{n,(N+1)p}} \frac{(b^\top u_i)^2 \sigma_i^2}{(\lambda^* - \sigma_i^2)^2},
    \end{align*}
    where $u_i$ are the columns of $U$. By \Cref{as: observable}, we know $n < (N+1)p$. Observe that
    \begin{align*}
        f(\lambda) &:= \sum_{i=1}^{n} \frac{(b^\top u_i)^2 \sigma_i^2}{(\lambda - \sigma_i^2)^2},
    \end{align*}
    is a strictly monotonically decreasing function when $\lambda > \sigma_1^2$, and converges to $0$ when $\lambda \to \infty$. This implies there is a unique $\lambda^*$ such that $f(\lambda^*) = \varepsilon^2$, which can be numerically solved for in various ways, such as bisection. Such methods can be initialized by setting the left boundary to $\lambda_{\ell} := \sigma_1^2$, corresponding to $f(\lambda_{\ell}) = \infty$, and the right boundary to a precomputable over-estimate $\lambda_r$ such that $f(\lambda_r) < \varepsilon^2$. As an example, one such crude over-estimate can be derived by observing
    \begin{align*}
        \sum_{i=1}^n \frac{(b^\top u_i)^2 \sigma_i^2}{(\lambda - \sigma_i^2)^2} &\leq \paren{\sum_{i=1}^n \paren{b^\top u_i}^2} \frac{\sigma_1^2}{\paren{\lambda - \sigma_1^2}^2} && \text{H\"older's inequality} \\
        &= \norm{b}^2 / \paren{c^2 \sigma_1^2} \leftarrow \varepsilon^2 && U \text{ orthogonal; set }\lambda = (1+c)\sigma_1^2 \\
        \implies \lambda_r &:= \paren{1 + \frac{\norm{b}}{\varepsilon \sigma_1}}\sigma_1^2.
    \end{align*}
    Therefore, bisection can solve for $\lambda^*$ up to a desired tolerance $\delta$ in at most $\ceil{\log_2\paren{\frac{\lambda_r - \lambda_\ell}{\delta}}} = \ceil{\log_2\paren{\frac{\norm{b}}{\varepsilon \sigma_1 \delta}}}$ iterations, which is linear bit-complexity over problem parameters. Since $f(\lambda)$ enjoys favorable regularity properties such as monotonicity, strict convexity, and smoothness on an open interval around $\lambda^*$, more advanced root-finding methods such as variants of Newton's method or the secant method can be employed for superlinear convergence.
    
    \item Now we consider the case where $\lambda^* = \sigma_1^2$. In this case, $\delta^*$ will no longer be unique, and will come in the form
    \begin{align*}
        \delta^* &= -V (\sigma_1^2 I-\Sigma^\top \Sigma)^{\dagger} \Sigma^\top U^\top b + cv,
    \end{align*}
    where $^\dagger$ denotes the Moore-Penrose pseudoinverse, and $v$ is any unit vector lying in the null-space of $\paren{\sigma_1^2 I -\Sigma^2 }V^\top$, which is precisely characterized in this case by
    \begin{align*}
        \ker\paren{\paren{\sigma_1^2 I-\Sigma^2}V^\top} &= \Span\paren{\curly{v_i : \sigma_i^2 = \sigma_1^2}},
    \end{align*}
    with $v_i$ denoting the $i$th column of $V$. To find the appropriate scaling $c$, we observe
    \begin{align*}
        {\delta^*}^\top \delta^* &=  b^\top U \Sigma \paren{(\sigma_1^2 I - \Sigma^\top \Sigma)^{\dagger}}^2 \Sigma^\top U^\top b + c^2 \\
        &= \sum_{i:\sigma_i < \sigma_1} \frac{(b^\top u_i)^2 \sigma_i^2}{(\sigma_1^2 - \sigma_i^2)^2} =\varepsilon^2 \\
        c &= \sqrt{\varepsilon^2 - \sum_{i:\sigma_i < \sigma_1} \frac{(b^\top u_i)^2 \sigma_i^2}{(\sigma_1^2 - \sigma_i^2)^2}}.
    \end{align*}
    Combining our precise characterization of $\ker\paren{\paren{\sigma_1^2 I-\Sigma^\top\Sigma}V^\top}$ using the columns of $V$, and the formula for $c$, we can extract an optimal perturbation vector $\delta^*$. Therefore, we have demonstrated that (P), as well as extracting its optimal solution, can be solved to arbitrary precision.

\end{itemize}
\section{General Statements and Proofs for Section \ref{sec: Kalman Bounds}} \label{ss: kf general statements and proofs}

\subsection{Proof of Theorem \ref{thm:generalUBLB}} \label{ss:proof of general UBLB}

\textbf{Theorem \ref{thm:generalUBLB}}\;\;\emph{Given any $L \in \R^{p \times n}$, we have the following lower bound on $\AR(L) - \SR(L)$:
\begin{align*}
    \AR(L) - \SR(L) &\geq 2\varepsilon\; \Ex_{x,w}\brac{\norm{L^\top (x - Ly)}_2}  + \varepsilon^2 \lambda_{\min} (L^\top L),
\end{align*}
and a corresponding upper bound
\begin{equation*}
    \AR(L) - \SR(L)  \leq 2\varepsilon\; \Ex_{x,w}\brac{\norm{L^\top (x - Ly)}_2} + \varepsilon^2 \lambda_{\max}(L^\top L),
\end{equation*}
where $\lambda_{\min}(L^\top L)$ and $\lambda_{\max}(L^\top L)$ are the minimum and maximum eigenvalues of $L^\top L$, respectively.
}

\textbf{Proof:} First recall the definitions of SR and AR:
\begin{align*}
    \SR(L) &= \Ex\brac{\norm{x - Ly}_2^2} \\
    \AR(L) &= \Ex\brac{\max_{\norm{\delta}_2 \leq \varepsilon}\norm{x - L(y+\delta)}_2^2}.
\end{align*}
Given fixed $x, y$, let us define the quantity $d(L) := x - Ly$. Writing out the inner maximization of AR we have:
\begin{align*}
    \max_{\norm{\delta}_2 \leq \varepsilon}\norm{x - L(y+\delta)}_2^2 &= \max_{\norm{\delta}_2 \leq \varepsilon}\norm{d(L) - L\delta}_2^2.
\end{align*}
Observe that this is equivalent to the problem
\begin{align*}
    \minimize_{s} \quad& s \tag{P1}\\
    \subjectto \quad& s - \norm{d(L) - L\delta}_2^2 \geq 0 \text{ for all } \delta^\top \delta \leq \varepsilon^2.
\end{align*}

We now recall the S-lemma for quadratic functions.
\begin{lemma}[S-lemma] \label{lemma: S-lemma}
    Given quadratic functions $p(x), q(x) : \R^n \to \R$, suppose there exists $x$ such that $p(x) > 0$. Then,
    \[
    p(x) \geq 0 \implies q(x) \geq 0 \text{ for all }x
    \]
    if and only if
    \[
    \exists t \geq 0 \text{ such that }q(x) \geq tp(x)  \text{ for all }x.
    \]
\end{lemma}

Using this lemma, we set $p(\delta) = \varepsilon^2 - \delta^\top \delta$, $q(\delta) = s - \norm{d(L) - L\delta}_2^2 \geq 0$. We observe that trivially, there exists $\delta = 0$ such that $p(\delta) > 0$. Now given feasible $s$ for (P1), we observe that by our constraints, any $\delta$ such that $p(\delta) \geq 0$ immediately implies $q(\delta) \geq 0$. By the S-lemma, this is equivalent to the existence of some $t \geq 0$ such that
\[
q(\delta) - tp(\delta) = s - \norm{d(L) - L\delta}_2^2 - t(\varepsilon^2 - \delta^\top \delta) \geq 0
\]
for all $\delta$. Therefore, we can re-write the optimization problem (P1) into
\begin{align*}  
    \minimize_{s} \quad& s \tag{P2}\\
    \subjectto \quad& \exists t \geq 0 \text{ s.t. }s - \norm{d(L) - L\delta}_2^2 - t(\varepsilon^2 - \delta^\top \delta) \geq 0 \text{ for all }\delta.
\end{align*}
Re-arranging the terms in the quadratic expression, we get:
\begin{align*}
    s - \norm{d(L) - L\delta}_2^2 - t(\varepsilon^2 - \delta^\top \delta) &= s - \paren{\delta^\top L^\top L \delta - 2d(L)^\top L \delta + \norm{d(L)}_2^2} - t(\varepsilon^2 - \delta^\top \delta) \\
    &= \delta^\top \paren{tI - L^\top L}\delta + 2d(L)^\top L\delta + \paren{s - t\varepsilon^2 - \norm{d(L)}_2^2}.
\end{align*}
Now we recall a property of Schur complements.

\begin{lemma}[Schur Complement]
    Given $p(x) = x^\top P x + b^\top x + c$, we have
    \begin{align*}
    p(x) \geq 0 \;\forall\; x &\iff \bmat{P & b \\ b^\top & c} \succeq 0 \\
    &\iff P \succeq 0, \; c - b^\top P^\dagger b \geq 0.
    \end{align*}
\end{lemma}

Applying this to (P2), we see the constraints can be re-written
\begin{align*}
    &\exists t \geq 0 \text{ s.t. }s - \norm{d(L) - L\delta}_2^2 - t(\varepsilon^2 - \delta^\top \delta) \geq 0 \text{ for all }\delta \\
    \iff\quad& \exists t \geq 0, \; tI - L^\top L \succeq 0, \; s - t\varepsilon^2 - \norm{d(L)}_2^2 - d(L)^\top L (tI - L^\top L)^\dagger L^\top d(L) \geq 0 \\
    \iff\quad& \exists t\geq \lambda_{\max}(L^\top L),\;s - t\varepsilon^2 - \norm{d(L)}_2^2 - d(L)^\top L (tI - L^\top L)^\dagger L^\top d(L) \geq 0.
\end{align*}
Therefore, we get the optimization problem
\begin{align*}
    \minimize_{s,t} \quad& s \\
    \subjectto \quad& t\geq \lambda_{\max}(L^\top L)\\
    &s - t\varepsilon^2 - \norm{d(L)}_2^2 - d(L)^\top L (tI - L^\top L)^\dagger L^\top d(L) \geq 0.
\end{align*}
However, this is clearly equivalent and has the same optimal value as the following problem
\begin{align*}
    \minimize_{t} \quad& t\varepsilon^2 + \norm{d(L)}_2^2 + d(L)^\top L (tI - L^\top L)^\dagger L^\top d(L) \\
    \subjectto \quad& t\geq \lambda_{\max}(L^\top L).
\end{align*}
Notice that so far we are simply considering equivalent formulations to the original optimization. The ensuing step is where the lower and upper bounds \eqref{eq: s-lemma lower bound} and $\eqref{eq: s-lemma upper bound}$ arise. Recall the Neumann series, where since $t \geq \lambda_{\max}(L^\top L)$, we have
\begin{align*}
    (tI - L^\top L)^{-1} &= \frac{1}{t}\paren{I - \frac{1}{t}L^\top L }^{-1} \\
    &= \frac{1}{t}\paren{I + \frac{1}{t}L^\top L + \frac{1}{t^2}\paren{L^\top L}^2 + \cdots }.
\end{align*}
From the Neumann series, we see that we can upper and lower bound the inverse using geometric series of the largest and smallest eigenvalues of $L^\top L$, respectively,
\begin{align*}
    \frac{1}{t - \lambda_{\min}(L^\top L)} I \preceq (tI - L^\top L)^{-1} \preceq \frac{1}{t - \lambda_{\max}(L^\top L)} I
\end{align*}
From now on, we will deal with the inverse, since instead of the pseudo-inverse we can take the infimum of the above problem, which is bounded from below. Let us consider the lower bound first. The upper bound follows using the exact same analysis. We have
\begin{align*}
    t\varepsilon^2 + \norm{d(L)}_2^2 + d(L)^\top L (tI - L^\top L)^{-1} L^\top d(L) &\geq t\varepsilon^2 + \norm{d(L)}_2^2 + d(L)^\top L \paren{\frac{1}{t - \lambda_{\min}(L^\top L)} I } L^\top d(L) \\
    &= t\varepsilon^2 + \norm{d(L)}_2^2 + \frac{1}{t - \lambda_{\min}(L^\top L)}\norm{L^\top d(L)}_2^2.
\end{align*}
Therefore, we have
\begin{align*}
    \max_{\norm{\delta}_2 \leq \varepsilon}\norm{x - L(y+\delta)}_2^2 &\geq \norm{d(L)}_2^2 + \min_{t > \lambda_{\max}(L^\top L)} t\varepsilon^2  + \frac{1}{t - \lambda_{\min}(L^\top L)}\norm{L^\top d(L)}_2^2.
\end{align*}
We now make a second relaxation:
\begin{align*}
    \norm{d(L)}_2^2 + \min_{t > \lambda_{\max}(L^\top L)} t\varepsilon^2  + \frac{1}{t}\norm{L^\top d(L)}_2^2 &\geq \norm{d(L)}_2^2 + \min_{t \geq 0} t\varepsilon^2  + \frac{1}{t - \lambda_{\min}(L^\top L)}\norm{L^\top d(L)}_2^2 \\
    &= \norm{d(L)}_2^2 + 2\varepsilon \norm{L^\top d(L)}_2 + \varepsilon^2 \lambda_{\min}(L^\top L),
\end{align*}
which we get by deriving the unconstrained minimizer $t^* = \frac{\norm{L^\top d(L)}_2}{\varepsilon} + \lambda_{\min}(L^\top L)$. Now putting expectations on both sides of the inequality, we get
\begin{align*}
    \AR(L) &\geq \SR(L) + 2\varepsilon\;\Ex\brac{\norm{L^\top (x - Ly)}_2} + \varepsilon^2 \lambda_{\min}(L^\top L).
\end{align*}
\hfill$\blacksquare$



In the subsequent full statements of the corresponding results in the paper, we do not make \Cref{as:noisepd}, and instead consider general observable $(A,C)$, where $\rho(A) \leq 1$, and positive definite noise covariances $\Sigma_0, \Sigma_w, \Sigma_v \succ 0$.

\subsection{General Statement of Lemma \ref{lem:SR closed form}}
\textbf{Lemma~\ref{lem:SR closed form}.}\;\;\emph{The standard risk may be expressed as
    \begin{align*}
        \SR(L) &:= \Ex \brac{\norm{x_k - L Y_N}_2^2} \\
        &= \norm{\paren{A^k - L \calO_N}\Sigma_0^{1/2}}_F^2 + \norm{\paren{\Gamma_k - L \tau_N}\paren{I_N \otimes \Sigma_w}^{1/2}}_F^2 + \norm{L \paren{I_{N+1} \otimes \Sigma_v}^{1/2}}_F^2.
    \end{align*} 
}

\textit{Proof:} This follows simply by expanding the norm inside the expectation, and noticing that since $x_0$, $W_N$, $V_N$ are defined to be zero-mean Gaussian random vectors, their cross terms vanish. More precisely, we have
\begin{align*}
    \Ex \brac{\norm{x_k - L Y_N}_2^2} &= \Ex\brac{\tr\paren{(x_k - LY_N)(x_k - LY_N)^\top}}\\
    &= \Ex\bigg[\tr\bigg( \paren{A^k - L\calO_N}x_0x_0^\top \paren{A^k - L\calO_N}^\top + \paren{\Gamma_k - L\tau_N}W_NW_N^\top (\Gamma_k - L\tau_N)^\top \\
    &\quad + LV_NV_N^\top L^\top \bigg) \bigg] + 0 \\
    &= \norm{\paren{A^k - L \calO_N}\Sigma_0^{1/2}}_F^2 + \norm{\paren{\Gamma_k - L \tau_N}\paren{I_N \otimes \Sigma_w}^{1/2}}_F^2 + \norm{L \paren{I_{N+1} \otimes \Sigma_v}^{1/2}}_F^2.
\end{align*}
\hfill$\blacksquare$

\subsection{General Statement of Lemma \ref{thm:AR-SR gap lower bound}}
\textbf{Lemma~\ref{thm:AR-SR gap lower bound}.}\;\:\emph{For any $L \in \R^{n \times p(N+1)}$, the gap between $\AR(L)$ and $\SR(L)$ admits the following lower bound:
    \begin{align}\label{eq:AR-SR gap no exp}
         \AR(L) - \SR(L)
         &\geq  2\sqrt{\frac{2 }{\pi}} \frac{\varepsilon}{\sqrt{n}}  \tr\bigg(\bigg( L^\top \Big(S \Sigma_0 S^\top + T \paren{I_N \otimes\Sigma_w} T^\top \nonumber + L\paren{I_{N+1} \otimes \Sigma_v} L^\top\Big)L\bigg)^{1/2}\bigg) \\
         &\geq 2\sqrt{\frac{2 }{\pi}} \frac{\varepsilon}{\sqrt{n}} \sigma_{\min}(\Sigma_v)^{1/2} \norm{L}_F^2
    \end{align}
    where $S := A^k - L\calO_N$, $T := \Gamma_k - L\tau_N$.
}

\textit{Proof:} Applying the lower bound~\eqref{eq: s-lemma lower bound}, we have
 \begin{align*}
    \AR(L) &\geq \SR(L) + 2\varepsilon\;\Ex\brac{\norm{L^\top (x_k - LY_T)}_2}.
\end{align*}
 
Then to derive the lower bound \eqref{eq:AR-SR gap no exp}, we observe that the random vector  \begin{align*}
    z &= L^\top (x_k - LY_T) \\
    &= L^\top \paren{S x_0 -T W_T + LV_T)},
\end{align*}
is a zero-mean Gaussian with covariance
\begin{align*}
    \Sigma &= L^\top \paren{S\Sigma_0S^\top + T \Sigma_w T^\top + L \Sigma_v L^\top}L.
\end{align*}

We can also write $z = \Sigma^{1/2} w$ where $w \sim \calN(0,I)$.  Consider the diagonalization of $\Sigma^{1/2} = VSV^\top$. Then 
\begin{align*}
    \Ex \brac{\norm{z}_2} &= \Ex\brac{\norm {V S V^\top w}_2} \\
    &= \Ex\brac{\norm{\sum_i s_i w_i v_i }_2},
\end{align*}
where $v_i$ is the $i$th row of $V$ and $s_i$ is the $i$th singular value of $\Sigma^{1/2}$. We have that 
\begin{align*}
    \norm{\sum_i s_i w_i v_i }_2^2 = \sum_{i} s_i^2 w_i^2 v_i^\top v_i = \sum_{i} s_i^2 w_i^2
\end{align*}
We have by equivalence of norms, $\sqrt{\sum_{i=1}^n x_i^2} \geq n^{-1/2} \sum_{i=1}^n \abs{x_i}$. Therefore,
\begin{align*}
    \sqrt{\sum_{i=1}^n s_i^2 w_i^2} \geq \frac{1}{\sqrt{n}} \sum_{i=1}^n s_i \abs{w_i},
\end{align*}
and thus
\begin{align*}
    \Ex \brac{\norm{z}_2} &\geq \frac{1}{\sqrt n} \Ex \brac{\sum_{i=1}^n s_i |w|_i} = \frac{1}{\sqrt n} \Ex\brac{\abs{w}} \sum_{i=1}^n s_i 
\end{align*}
where $w \sim \calN(0,1)$. The quantity $\Ex\brac{\abs{w}}$ is the expected value of a folded standard normal, which is $\sqrt{\frac{2}{\pi}}$, while $\sum_{i=1}^n s_i  = \tr\paren{\Sigma^{1/2}}$. Putting this together, we have that
\begin{align*}
    \AR(L) - \SR(L)
         &\geq  2\sqrt{\frac{2 }{\pi}} \frac{\varepsilon}{\sqrt{n}}  \tr\bigg(\bigg( L^\top \Big(S \Sigma_0 S^\top + T \paren{I_N \otimes\Sigma_w} T^\top \nonumber + L\paren{I_{N+1} \otimes \Sigma_v} L^\top\Big)L\bigg)^{1/2}\bigg).
\end{align*}
From the above bound, we may now derive a cruder lower bound from which we can observe a dependence on the singular values of the observability grammian, $W_o(N)$. In particular, begin with the expression above, and note that the terms involving $\Sigma_0$ and $\Sigma_w$ are positive definite to achieve a lower bound in terms of $L$:
\begin{align*}
    \quad &2  \sqrt{\frac{2 }{\pi}} \frac{\varepsilon}{\sqrt{n}} \tr \left(\left(L\T \left(S \Sigma_0 S\T + T \paren{I_N \otimes \Sigma_w} T\T + L \paren{I_{N+1} \otimes \Sigma_v} L\T \right)L\right)^{1/2}\right) \\
    \geq \;&2\sqrt{\frac{2 }{\pi}} \frac{\varepsilon}{\sqrt{n}} \sigma_{\min}\left(\Sigma_v  \right)^{1/2} \tr \paren{\paren{L\T L L\T L}^{1/2}} \\
    = \;&2\sqrt{\frac{2 }{\pi}} \frac{\varepsilon}{\sqrt{n}} \sigma_{\min}\left(\Sigma_v  \right)^{1/2} \norm{L}_F^2. 
\end{align*} 
which completes the proof of inequality \eqref{eq:AR-SR gap no exp}. 
\hfill $\blacksquare$

Introducing additional notation to express the Kalman estimator will be helpful in subsequent sections. Let
\begin{align}\label{eq:lbnotation}
    \bar{\Sigma} &:= \bmat{\Sigma_0 & \\ & I_N \otimes \Sigma_w} \\ 
    \nonumber
    H &:=  \bmat{I &  & & & \\ A  & I & & & \\ \vdots && \ddots && \\ A^N & A^{N-1} & \hdots & I}\bar{\Sigma}^{1/2} \\
    \nonumber
    H_k &:= E_k^\top H =  \bmat{A^k& A^{k-1} & \dots & I & 0 & \dots & 0}\bar{\Sigma}^{1/2} \\
    \nonumber
    M &:= H^\top \paren{C^\top \otimes I} 
    = \bar{\Sigma}^{1/2} \bmat{ \calO_N^\top \\  \paren{\calZ \calO_N}^\top \\ \vdots \\ \paren{\calZ^N \calO_N}^\top}, 
\end{align}
where $\calZ \in \R^{p(N+1) \times p(N+1)}$ is a block downshift operator, with blocks of size $m$. With this notation, the Kalman estimator given in Lemma \ref{lem:kf} may be rewritten more compactly as
\[
    \hat{L}_k = H_k M\paren{M^\top M + \Sigma_v}^{-1}.
\]

\subsection{General Statement of Theorem \ref{thm:lb}}

\textbf{Theorem \ref{thm:lb}.}\;\:\emph{Suppose that $\hat{L}_k$ is the Kalman estimator from Lemma~\ref{lem:kf}.
    Then we have the following bound on the gap between $\AR$ and $\SR$. 
    \begin{align*}
        \AR(\hat{L}_k) - \SR(\hat{L}_k) \geq 2\sqrt{\frac{2 }{\pi}} \frac{\varepsilon}{\sqrt{n}}\sigma_{\min}\left(\Sigma_v  \right) \norm{C}_F^2\paren{ \frac{\sigma_{\min} \paren{A^k \Sigma_0 {A^k}^\top + \sum_{i=1}^k A^{k-i} \Sigma_w {A^{k-i}}^\top} }{(N+1) \norm{\bar{\Sigma}}_2^2 \norm{W_o(N)}_F + \norm{\Sigma_v}_2} }^2. 
    \end{align*}
}


\emph{Proof:}
We begin by writing the Kalman estimator using the notation defined in \eqref{eq:lbnotation}
\[L_k = H_k M\paren{M^\top M + I_{N+1} \otimes \Sigma_v}^{-1}.\]
Suppose the rank of $M$ is $m$. Then the singular value decomposition of $M$ can be taken to be 
\[ U \bmat{S & 0 \\ 0 & 0} V^\top = M\]
where $S = \diag\paren{\bmat{s_1 & s_2 & \dots s_{m}}}$ with $s_1 \geq s_2 \geq \dots \geq s_m \geq 0$, while $U\in\R^{n(N+1)\times n(N+1)}$ and $V\in\R^{p(N+1)\times p(N+1)}$. We can now lower bound the Frobenius norm of $L_k$ as follows.
\begin{align*}
    \norm{L_k}_F^2 \geq \norm{H_k M}_F^2 \sigma_{\min}\curly{\paren{M^\top M + \Sigma_v I}^{-1}}^2.
\end{align*}
Note that $\sigma_{\min}\curly{\paren{M^\top M + \Sigma_v I}^{-1}} \geq \sigma_{\min}\curly{\paren{M^\top M + \norm{\Sigma_v}_2 I}^{-1}}$. Therefore 
\begin{equation}\label{eq:fnlb}
\begin{aligned}
    \norm{L_k}_F^2  &\geq \sigma_{\min} \curly{\paren{\bmat{S^2 & \\ & 0} + \norm{\Sigma_v}_2 I}^{-2}} \norm{H_k M}_F^2 \\
    &= \sigma_{\min} \curly{\paren{\bmat{S^2 & \\ &0} + \norm{\Sigma_v} I}^{-2}} \norm{H_kH^\top \paren{C^\top \otimes I}}_F^2 \\
    = \sigma_{\min} \curly{\paren{\bmat{S^2 & \\ &0} + \norm{\Sigma_v}_2 I}^{-2}} & \norm{\bmat{A^k \Sigma_0 C^\top & \ldots & A^k \Sigma_0 \paren{A^{N}}^\top C^\top+ \sum_{i=1}^k A^{k-i} \Sigma_w \paren{A^{N-i}}^\top C^\top }}_F^2
\end{aligned}
\end{equation}

Now observe that 
\begin{equation}\label{eq:smlb}
    \sigma_{\min} \curly{\paren{\bmat{S^2 & \\ & 0} + \norm{\Sigma_v} I}^{-2}} = \frac{1}{\paren{s_1^2 + \norm{\Sigma_v}_2}^2}
\end{equation}    
Also note that $s_1 = \norm{M}_2 \leq \norm{\bar \Sigma^{1/2}}_2 \norm{ \bar \Sigma^{-1/2} M}_F$. We have that 
\begin{align*}
    \norm{\bar \Sigma^{-1/2} M}_F^2 = \norm{ \bmat{ \calO_N^\top \\  \paren{\calZ \calO_N}^\top \\ \vdots \\ \paren{\calZ^N \calO_N}^\top}}_F^2 \leq \sum_{i=0}^N \norm{\calZ^i \calO_N}_F^2 \leq (N+1) \norm{\calO_N}_F^2.  
\end{align*}
Then 
\begin{align}\label{eq:s1lb}
s_1 \leq \norm{\bar \Sigma^{1/2}}_2 \sqrt{N+1} \norm{\calO_N}_F
\end{align}


When $k\geq 0$, we have 
\begin{align*}
    \norm{\bmat{A^k \Sigma_0 C^\top & \ldots & A^k \Sigma_0 \paren{A^{N}}^\top C^\top+ \sum_{i=1}^k A^{k-i} \Sigma_w \paren{A^{T-i}}^\top C^\top }}_F^2 \\
    \geq \norm{\paren{A^k \Sigma_0 \paren{A^k}^\top + \sum_{i=1}^k A^{k-i} \Sigma_w \paren{A^{k-i}}^\top}C^\top}_F^2 \\
    \geq\sigma_{\min} \paren{A^k \Sigma_0 \paren{A^k}^\top + \sum_{i=1}^k A^{k-i} \Sigma_w \paren{A^{k-i}}^\top}^2 \norm{C}_F^2. 
\end{align*}
In conjunction with \eqref{eq:fnlb}, \eqref{eq:smlb}, and \eqref{eq:s1lb}, this leads to \eqref{eq:lb}. \hfill $\blacksquare$


\subsection{Proof of Lemma \ref{thm:general upper bound}}
\textbf{Lemma \ref{thm:general upper bound}.}\;\:\emph{For any $L \in \R^{n \times p(N+1)}$, the following bound holds
    \begin{align*}
        \AR(L) -  \SR(L) \leq 2\varepsilon  \norm{L}_2 \norm{\Sigma^{1/2}}_F + \varepsilon^2 \norm{L}_2^2
    \end{align*}
    where $\Sigma^{1/2}$ is the symmetric square root of the covariance of $x_k - LY_N$. }

\textit{Proof:} By Theorem \ref{thm:generalUBLB}, 
\begin{align*}
    \AR(L) - \SR(L) \leq 2 \varepsilon  \Ex \brac{\norm{L(x_t - LY_N)}_2} + \varepsilon^2 \lambda_{\max}\paren{L^\top L} \leq 2 \varepsilon \norm{L}_2  \Ex \brac{\norm{x_t - LY_N}_2} + \varepsilon^2 \norm{L}_2^2
\end{align*}
We can upper bound $\Ex\brac{\norm{x_t - LY_N}_2}$ by bounding the expectation of the euclidean norm of a normal random variable. In particular, let $w$ be a $n$ dimensional standard normal random variable, so that 
$\Ex\brac{\norm{x_t - LY_N}_2} =\Ex\brac{\norm{ \Sigma^{1/2} w}_2}$, where $\Sigma$ is defined as the covariance of $x_k - LY_N$, and $\Sigma^{1/2}$ is its symmetric square root. 
Let $US^{1/2} U\T := \Sigma^{1/2}$ be the eigenvalue decomposition of $\Sigma^{1/2}$ so that
\begin{align*}
    \norm{\Sigma^{1/2} z}_2 &= \norm{U S^{1/2} U\T w}_2
\end{align*}
Now define $z = U\T w$. We have that $z \sim N(0, I)$. Then the above quantity equals $\sqrt{\norm{U S^{1/2} w}_2^2}$. Jensen's inequality tells us that
\begin{align*}
    \Ex \brac{\sqrt{\norm{US^{1/2}  w}_2^2}} \leq \sqrt{\Ex \brac{\norm{U S^{1/2} w}_2^2}} = \sqrt{\Ex \brac{w\T S^{1/2} w}} = \norm{S^{1/2}}_F = \norm{\Sigma^{1/2}}_F,
\end{align*}
from which the theorem follows. \hfill $\blacksquare$

\subsection{General Statement of Theorem \ref{thm:ub}}

\textbf{Theorem~\ref{thm:ub}.}\;\:\emph{Suppose that $\hatL_k$ is the Kalman state estimator given by Lemma \ref{lem:kf}. Then the gap between $\AR(\hatL_k)$ and $\SR(\hatL_k)$ is upper bounded by
\begin{align*}
    \AR(\hatL_k) - \SR(\hatL_k) &\geq \varepsilon  \paren{\frac{{\norm{A^k \Sigma_0 \paren{A^k}^\top + \sum_{i=1}^k A^{k-i} \Sigma_w \paren{A^{k-i}}^\top_2}_2}}{\sigma_{\min}(W_o(N))^{1/2} \sigma_{\min} \paren{\bar\Sigma}}} \\
    & \quad \times \left(2 \sqrt{n}
        \paren{\norm{\bar{\Sigma}}_2 + \paren{\frac{\sqrt{\norm{\Sigma_v}_2}}{\sigma_{\min}(W_o(N))^{1/2} \sigma_{\min} \paren{\bar\Sigma}}}^2 }^{1/2}  
    +\varepsilon \paren{ \frac{1}{\sigma_{\min}(W_o(N))^{1/2} \sigma_{\min} \paren{\bar\Sigma}}}\right).
\end{align*}
Furthermore, if $\lambda_{\min}(W_o(N))^{1/2} \geq \sigma_v/\sigma_{\min}\paren{\bar\Sigma}$, and defining $\kappa = \frac{\lambda_{\min}(W_o(N))^{1/2} \sigma_{\min} \paren{\bar\Sigma}}{\lambda_{\min}(W_o(N)) \sigma_{\min} \paren{\bar\Sigma}^2 + \sigma_v}$, we get the bound
\begin{align*}
    \AR(\hat{L}_k) - \SR(\hat{L}_k) &\leq \varepsilon \paren{\kappa \sqrt{\norm{A^k \Sigma_0 \paren{A^k}^\top + \sum_{i=1}^k A^{k-i} \Sigma_w \paren{A^{k-i}}^\top_2}_2}  } \\
    &\quad \times \paren{2\sqrt{n}\paren{\sigma_{\max}(\bar{\Sigma})^2 + \sigma_{\min}(\Sigma_v) \kappa^2}^{1/2} + \varepsilon \kappa }
\end{align*}
}

\textit{Proof:} By Lemma \ref{thm:general upper bound}, upper bounding the gap between $\AR(L_k)$ and $\SR(L_k)$ reduces to upper bounding $\norm{\Sigma^{1/2}}_F$ and $\norm{L_k}_2$. First consider $\norm{\Sigma^{1/2}}_F$. Equivalence of norms tells us that
\begin{align}\label{eq:norm equivalence}
    \norm{\Sigma^{1/2}}_F \leq \sqrt{n} \norm{\Sigma^{1/2}}_2 = \sqrt{n}\norm{\Sigma}_2^{1/2}.
\end{align}
Recalling the notation defined in \eqref{eq:lbnotation}, the Kalman filter may be expressed as $L_k = H_k M \paren{M^\top M + \Sigma_v}^{-1}$. We may also  express $\Sigma$ in terms of this notation: $\Sigma = \paren{H_k - L_k M} \bar{\Sigma} \paren{H_k - L_k M}^\top + L_k \paren{I \otimes \Sigma_v} L_k\T$. To upper bound the spectral radius of this, we can leverage triangle inequality and submultiplicativity
\begin{align*}
    \norm{\Sigma}_2
    &\leq  \norm{\paren{H_k - L_k M} \bar{\Sigma} \paren{H_k - L_kM}^\top}_2 + \norm{L_k \paren{I \otimes \Sigma_v} L_k\T}_2 \\
    &\leq  \norm{\bar{\Sigma}}_2 \norm{H_k - L_kM}_2^2 + \norm{\Sigma_v}_2 \norm{L_k}_2^2.
\end{align*}
Note that $H_k - L_kM = H_k - H_k M\paren{M^\top M + \Sigma_v}^{-1} M = H_k \paren{I - M \paren{M^\top M + \Sigma_v}^{-1} M^\top}$. Then by submultiplicativity, 
\begin{align*}
    \norm{H_k \paren{I - M \paren{M^\top M + \Sigma_v}^{-1} M^\top}}_2 \leq \norm{H_k}_2 \norm{I - M \paren{M^\top M + \Sigma_v}^{-1} M^\top}_2 \leq \norm{H_k}_2.
\end{align*}
We can further upper bound $\norm{H_k}_2$ in terms of system properties. In particular, we have
\begin{align*}
    \norm{H_k}_2 =  \sqrt{\norm{H_k}_2^2} = \sqrt{\norm{H_k H_k^\top}_2} = \sqrt{\norm{A^k \Sigma_0 \paren{A^k}^\top + \sum_{i=1}^k A^{k-i} \Sigma_w \paren{A^{k-i}}^\top_2}_2}.
\end{align*}

Thus 
\begin{align}\label{eq:sigmabound}
    \norm{\Sigma}_2 \leq \norm{\bar{\Sigma}}_2 \norm{H_k}_2^2 + \norm{\Sigma_v}_2 \norm{L}_2^2\leq\norm{\bar{\Sigma}}_2 \norm{A^k \Sigma_0 \paren{A^k}^\top + \sum_{i=1}^k A^{k-i} \Sigma_w \paren{A^{k-i}}^\top_2}_2 + \norm{\Sigma_v}_2 \norm{L}_2^2.
\end{align}

Next we obtain a bound on $\norm{L_k}_2$. As in the proof of Theorem \ref{thm:lb}, we will assign $m: = \rank(M)$ and take the singular value decomposition of $M$ to be $U \bmat{S & 0 \\ 0 & 0} V^\top$ where $S = \diag\paren{\bmat{s_1 & \dots & s_m}}$ with $s_1 \geq s_2 \geq \dots \geq s_m \geq 0$, while $U \in\R^{n(N+1)\times n(N+1)}$ and $V \in \R^{p(N+1)\times p(N+1)}$. 
\begin{align*}
    \norm{L_k}_2 &= \norm{H_k M\paren{M^\top M + \Sigma_v}^{-1}}_2 \leq \norm{H_k}_2 \norm{M\paren{M^\top M + \sigma_{\min}\paren{\Sigma_v}}^{-1}}_2 \\
    &= \norm{H_k}_2\norm{U \bmat{S &  \\  & 0} V^\top \paren{V \paren{\bmat{S^2 & \\ & 0} + \sigma_{\min}(\Sigma_v)} V^\top}^{-1}} \\
    &= \norm{H_k}_2\norm{U \bmat{S & \\ & 0} \paren{\bmat{ S^2 & \\ & 0} + \sigma_{\min}(\Sigma_v)}^{-1}V^\top}_2 \\
        &= \norm{H_k}_2\norm{S (S^2 + \sigma_{\min}(\Sigma_v))^{-1}}_2 \\
        &\leq \norm{H_k}_2\max_{1\leq k \leq m}\frac{s_k}{s_k^2 + \sigma_{\min}(\Sigma_v)}.
\end{align*}

A simple bound on the last maximization would be
\begin{align}\label{eq: s_k upper bound}
    \max_{1\leq k \leq m}\frac{s_k}{s_k^2 + \sigma_{\min}(\Sigma_v)} &\leq \max_{1\leq k \leq m}\frac{s_k}{s_k^2} \leq \frac{1}{s_m}
\end{align}
Note that $s_m \geq \lambda_{\min}(W_o(N))^{1/2} \sigma_{\min} \paren{\bar\Sigma}$, 
so $\frac{1}{s_m} \leq \frac{1}{\lambda_{\min}(W_o(N))^{1/2} \sigma_{\min} \paren{\bar\Sigma}}$. Then 
\begin{align}\label{eq:Lbound}
    \norm{L_k}_2 \leq  \frac{\norm{H_k}_2}{\lambda_{\min}(W_o(N))^{1/2} \sigma_{\min} \paren{\bar\Sigma}} \leq \frac{\sqrt{\norm{A^k \Sigma_0 \paren{A^k}^\top + \sum_{i=1}^k A^{k-i} \Sigma_w \paren{A^{k-i}}^\top_2}_2}}{\lambda_{\min}(W_o(N))^{1/2} \sigma_{\min} \paren{\bar\Sigma}}.
\end{align}

Then the first half of the theorem follows by combining the result of Lemma \ref{thm:general upper bound} with \eqref{eq:norm equivalence},  \eqref{eq:sigmabound} and \eqref{eq:Lbound}.

However, if $\lambda_{\min}(W_o(N))^{1/2} \sigma_{\min} \paren{\bar\Sigma} \geq \sigma_v$, then the maximum \eqref{eq: s_k upper bound} is attained at
\begin{align*}
    \max_{1\leq k \leq m}\frac{s_k}{s_k^2 + \sigma_{\min}(\Sigma_v)} &= \frac{s_m}{s_m^2 + \sigma_v} \\
    &\leq \frac{\lambda_{\min}(W_o(N))^{1/2} \sigma_{\min} \paren{\bar\Sigma}}{\lambda_{\min}(W_o(N)) \sigma_{\min} \paren{\bar\Sigma}^2 + \sigma_v}.
\end{align*}
Therefore, when $\lambda_{\min}(W_o(N))^{1/2} \geq \sigma_v/\sigma_{\min} \paren{\bar\Sigma}$, we have the more precise bound
\begin{align*}
    \norm{L_k}_2 &\leq  \norm{H_k}_2 \frac{\lambda_{\min}(W_o(N))^{1/2} \sigma_{\min} \paren{\bar\Sigma}}{\lambda_{\min}(W_o(N)) \sigma_{\min} \paren{\bar\Sigma}^2 + \sigma_v} \\
    &\leq \sqrt{\norm{A^k \Sigma_0 \paren{A^k}^\top + \sum_{i=1}^k A^{k-i} \Sigma_w \paren{A^{k-i}}^\top_2}_2} 
    \frac{\lambda_{\min}(W_o(N))^{1/2} \sigma_{\min} \paren{\bar\Sigma}}{\lambda_{\min}(W_o(N)) \sigma_{\min} \paren{\bar\Sigma}^2 + \sigma_v},
\end{align*}
which leads to the second half of the theorem.


\hfill $\blacksquare$

\end{document}